\documentclass[superscriptaddress,prl, amsmath,amssymb,reprint,showkeys,showpacs]{revtex4-1}
\usepackage{amssymb,amsmath}
\usepackage{graphicx}
\usepackage{subfigure}
\usepackage[english]{babel}
\usepackage{float}
\usepackage{color}
\usepackage{booktabs}
\usepackage{blindtext}
\usepackage{comment}
\usepackage[symbol]{footmisc}

\usepackage{tikz,pgf}

\usepackage{hyperref}
\hypersetup{
    colorlinks=true,
    linkcolor=blue,
    citecolor=blue,
    filecolor=cyan,      
    urlcolor=cyan,
}
  \usepackage{tikz,pgf}
\newcommand{\mel}[3]{\langle#3|#2|#1\rangle}
 \newcommand{\ket}[1]{\left|#1\right\rangle}
 \newcommand{\bra}[1]{\left\langle#1\right|}
 
 \DeclareMathOperator{\Tr}{Tr}
\begin{document}

\title{Deterministic Generation of Large Fock States}

\author{M. Uria}
\affiliation{Departamento de F\'{\i}sica and Millennium Institute for Research in Optics (MIRO),  Facultad
de Ciencias F\'isicas y Matem\'aticas, Universidad de Chile,
Santiago, Chile}

\author{P. Solano}
\affiliation{Department of Physics, MIT-Harvard Center for Ultracold Atoms, and Research Laboratory of Electronics, Massachusetts Institute of Technology, Cambridge, Massachusetts 02139, USA}
\affiliation{Departamento de F\'isica, Facultad de Ciencias Físicas y Matemáticas, Universidad de Concepci\'on, Concepci\'on, Chile}

\author{C. Hermann-Avigliano}
\affiliation{Departamento de F\'{\i}sica and Millennium Institute for Research in Optics (MIRO),  Facultad
de Ciencias F\'isicas y Matem\'aticas, Universidad de Chile,
Santiago, Chile}

\thanks{Corresponding author: carla.hermann@uchile.cl}

\begin{abstract} 
We present a protocol to deterministically prepare the electromagnetic field in a large photon number state. The field starts in a coherent state and, through resonant interaction with one or few two-level systems, it evolves into a coherently displaced Fock state,  without any post-selection. We show the feasibility of the scheme under realistic parameters. The presented method opens a door to reach Fock states, with $n\sim100$ and optimal fidelities above $70$\%, blurring the line between macroscopic and quantum states of the field. 
\end{abstract}

\maketitle

{\it Introduction.---} Fock states are quantum states of the electromagnetic field with a well defined number of excitations. Such states are of significant theoretical and experimental interest, with applications ranging from protocols for quantum information to quantum metrology~\cite{HarocheBook}, where the quantum properties of the field allow for sensitivities greater than the achievable with classical light~\cite{Giovannetti_2011}. All these applications benefit from a fast and efficient generation of Fock states with a large number of photons, a long standing goal for the quantum optics community~\cite{varcoe2000,Brattke2001,Bertet2002,Waks2006,2008Natur.454..310H,Chu2018,Tiedau2019,Yanagimoto2019,DellAnno2006}.

There are several theoretical proposals and experimental implementations for generating Fock states across different platforms, such as acoustic waves in resonators~\cite{Chu2018}, photonic waveguides~\cite{GonzalesTudela2015,GonzalesTudela2017,GonzalesTudela2020} and superconducting circuits \cite{Leek2010,Premaratne2017,gu2017,hofheinz2009}. In the context of cavity quantum electrodynamics (CQED), Fock states can be generated by injecting one quanta at a time into a cavity field~\cite{Vogel1993,2008Natur.454..310H}, by resonantly interacting a jet of atoms passing through the cavity leaving the field in a upper-bounded steady Fock state~\cite{Meystre:88,PhysRevLett.82.3795}, or by realizing quantum non-demolition measurements progressively projecting the field into a Fock state~\cite{2007Natur.448..889G,Samuel2008Nature,2011Natur.477...73S,zhou2012PRL,Geremia2006}. 

State of the art experiments generate Fock states with either low photon number ($n\sim 7-15$)~\cite{zhou2012PRL,Wang2008,hofheinz2008}, low fidelity at large photon numbers ($F>80\%$ for $n\leq$ 4 and $F<50\%$ for $n\geq$ 4)~\cite{Samuel2008Nature}, or low probability of success after long convergence times ($\sim80\%$ after $\sim 20$~ms)~\cite{2011Natur.477...73S}, evidencing the difficulty of the problem and the efforts made to generate arbitrarily large number states.

\begin{figure}
  \includegraphics[width=.4\textwidth]{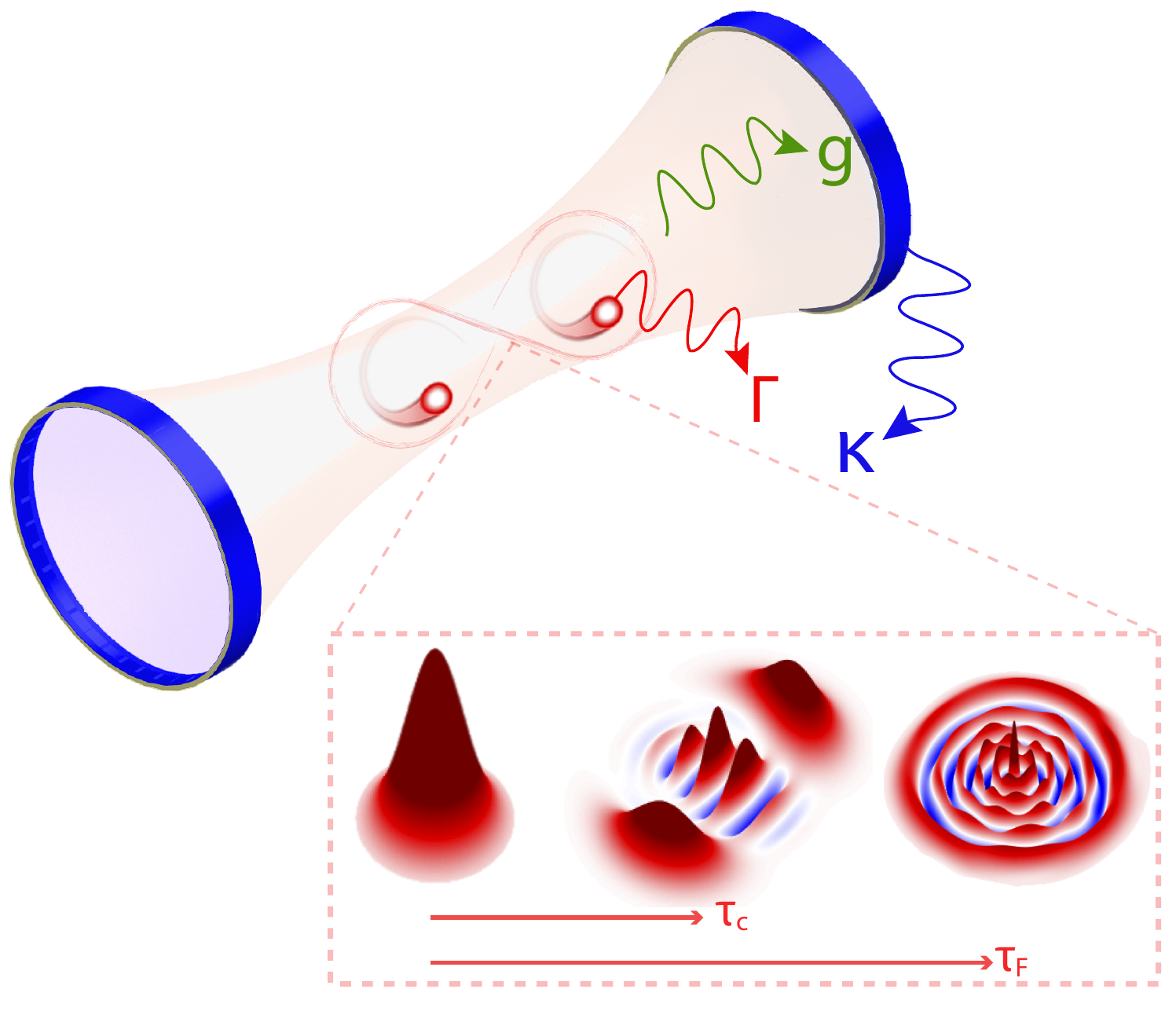}
 \caption{Schematic of the proposal. One or two (entangled) two-level atoms interact with a coherent state trapped in a cavity. The field, represented here by its Wigner function, evolves from a coherent state to a macroscopic superposition (after an interaction time $\tau_C$) and then into a Fock-like state (after an interaction time $\tau_{\text{F}}$). The time scale of the evolution is set by coupling strength $g$. The final state of the field approaches to a Fock-like state, despite the cavity and atomic decay, given by the rates $\kappa$ and $\Gamma$.}\label{fg:setup}
\end{figure}

In this Letter we propose a protocol to deterministicaly generate large photon number states with significantly large fidelities, depicted in Fig. \ref{fg:setup} with a CQED example. We consider a two-level atom that resonantly interacts with a coherent state. We show that for particular interaction times, the field evolves into a Fock-like state, slightly displaced in phase space. In the absence of decoherence, this protocol allows for fidelities above $70\%$ for $n\sim100$. This process can be sped up by simultaneously interacting two or more entangled atoms with the field. Under realistic losses, this scheme could generate Fock states as large as $n=50$ with a fidelity of $58\%$ in a CQED system ~\cite{Haroche2013Rev} and a Fock states with $n=100$ with a fidelity above $60\%$ considering the state of the art in circuit-QED \cite{Campagne2019,Kjaergaard2019,Landig2019}. Finally, we discuss the main characteristics and results of the protocol, and give an outlook of some open questions and future possibilities.

{\it Theoretical model.---} The interaction of an atom with the electromagnetic field inside a cavity is well described by the Jaynes-Cummings Hamiltonian \cite{JCM1963}
\begin{equation}
\label{EQ:jcm}
\hat{H}=\frac{\hbar\omega_0}{2}\hat{\sigma}_z+\hbar\omega_c \hat{a}^\dagger \hat{a} + \hbar g (\hat{a}\hat{\sigma}_++ \hat{a}^\dagger\hat{\sigma}_-),
\end{equation}
where $\omega_0$ and $\omega_c$ are the atomic and field frequencies, $g=\Omega_0/2$ is the coupling frequency, $\hat{a}$ and $\hat{a}^\dagger$ are the field operators, and $\hat{\sigma}_+$ and $\hat{\sigma}_-$ the raising and lowering atomic operators. The evolution under resonant interaction ($\omega_c=\omega_0$) of the atom-field compound state $\rho$ is determined by the master equation \cite{HarocheBook}
\begin{align}
 \label{eq:master}
 \hbar i \frac{d \rho(t)}{dt}&=[H_{\text{int}},\rho(t)]\nonumber\\ &-\frac{\kappa}{2}(n_{\text{th}}+1)\left(\hat{a}^{\dagger} \hat{a} \rho(t) +\rho(t) a^\dagger a-2a\rho(t) a^\dagger \right) \nonumber\\
 &- \frac{\kappa}{2}n_{\text{th}}\left(a a^{\dagger}  \rho(t) +\rho(t)  a a^\dagger-2a^\dagger \rho(t) a\right) \nonumber\\
 &-\frac{\Gamma}{2}(n_{\text{th}}+1)\left(\hat{\sigma}_+ \hat{\sigma}_- \rho(t) +\rho(t) \hat{\sigma}_+ \hat{\sigma}_--2\hat{\sigma}_-\rho(t) \hat{\sigma}_+ \right) \nonumber\\
 &- \frac{\Gamma}{2}n_{\text{th}}\left(\hat{\sigma}_- \hat{\sigma}_+  \rho(t) +\rho(t)  \hat{\sigma}_- \hat{\sigma}_+-2\hat{\sigma}_+ \rho(t) \hat{\sigma}_-\right)\,,
 \end{align}
where $\hat{H}_{\text{int}}=\hbar g (\hat{a}\hat{\sigma}_++ \hat{a}^\dagger\hat{\sigma}_-)$ is the Hamiltonian in the interaction representation, $\kappa$ and $\Gamma$ are the cavity and the atomic decay rates respectively, and $n_{\text{th}}$ is the average number of thermal photons.

We assume that the field is initialized in a coherent state of amplitude $\alpha$ and the state of the system is initially separable, meaning $\rho(0)=\ket{\psi(0)}\bra{\psi(0)}$ with $\ket{\psi(0)}=\ket{\phi_{\text{at}}}\ket{\alpha}$. 

The compound state initially factorizable, generally evolves into an entangled state. For a particular evolution time $\tau_{\text{C}}=2\pi\sqrt{\bar{n}}/\Omega_0$, with $\bar{n}=\vert\alpha\vert^2$ the average number of photons, the atom and the field get disentangled again and the field is found in a 
a cat-like state (see Fig. \ref{fg:setup}) \cite{Gea-Banacloche1990prl,Gea-Banacloche1991pra}. Previous works have studied this system within such a time regime \cite{Saavedra1997,Hermann2015pra}, nonetheless the exact evolution of the field at longer times has remain unexplored. 
 
If we let the system interact for longer times we see that the Wigner function of the field temporarily evolves into a distribution that, at specific times $t=\tau_{\text{F}}$, resembles that of a Fock state~\cite{SM} (see Fig. \ref{fg:setup}), but slightly displaced in phase space \cite{knight1990}. At $t=\tau_{\text{F}}$, the field and the atom become almost disentangled again, producing a field state that is nearly pure (purity $\approx80\%$)~\cite{SM}. By controlling the interaction time and injecting the proper field amplitude and phase to correct for the remnant coherent displacement, one can deterministically obtain a target Fock state. 

The more commonly used figure of merit to quantify how close is the generated state 
$\rho_{f}(t)=\text{Tr}_{\text{at}}\left[\rho(t)\right]$ to an ideal Fock state with $n$ photons ($\rho_n$) is the fidelity $F(\rho_n,\rho_f(t))=[Tr(\sqrt{\sqrt{\rho_n}\rho_f(t)\sqrt{\rho_n}})]^2$ \cite{1994JMO}. Since the fidelity is not a proper metric, we characterize how similar both states are by calculating the function $1-\delta(\rho_f(t),\rho_n)$, where $\delta(\rho_f(t),\rho_n)=\frac{1}{2}Tr(\vert \rho_f(t)-\rho_n\vert)$ is the trace distance \cite{PhysRevA.71.062310}. Although we use one minus the distance trace for our calculations, we present our results in terms of the fidelity to provide a common-ground comparison with previous works.

We numerically calculate the evolution of the field state $\rho_f(t)$ under Eq. (\ref{eq:master}) and search for an optimum time $t=\tau_{\text{F}}$ that maximizes the value of $1-\delta\left(\rho_f(t),\rho_n\right)$ for a target Fock state $\rho_n$. Because the field state evolves to something close to a Fock state but with a small displacement $D(\beta)$($=\text{Exp}\left[\beta a^{\dag}-\beta^* a\right]$), we applied a coherent displacement $-\beta$ after the field interacted with the atom \cite{SayrinThesis,Samuel2008Nature,hofheinz2009,Penasa2016}. We perform a numerical evaluation of the function $1-\delta\left(\rho_f(t,\beta),\rho_n\right)$ and optimize it over two parameters, namely $t$ and $\beta$, obtaining the optimal interaction time $\tau_{\text{F}}$ and its corresponding optimum coherent displacement $\beta_{\text{F}}$. 

 \begin{figure}[h]
  \includegraphics[width=0.48\textwidth]{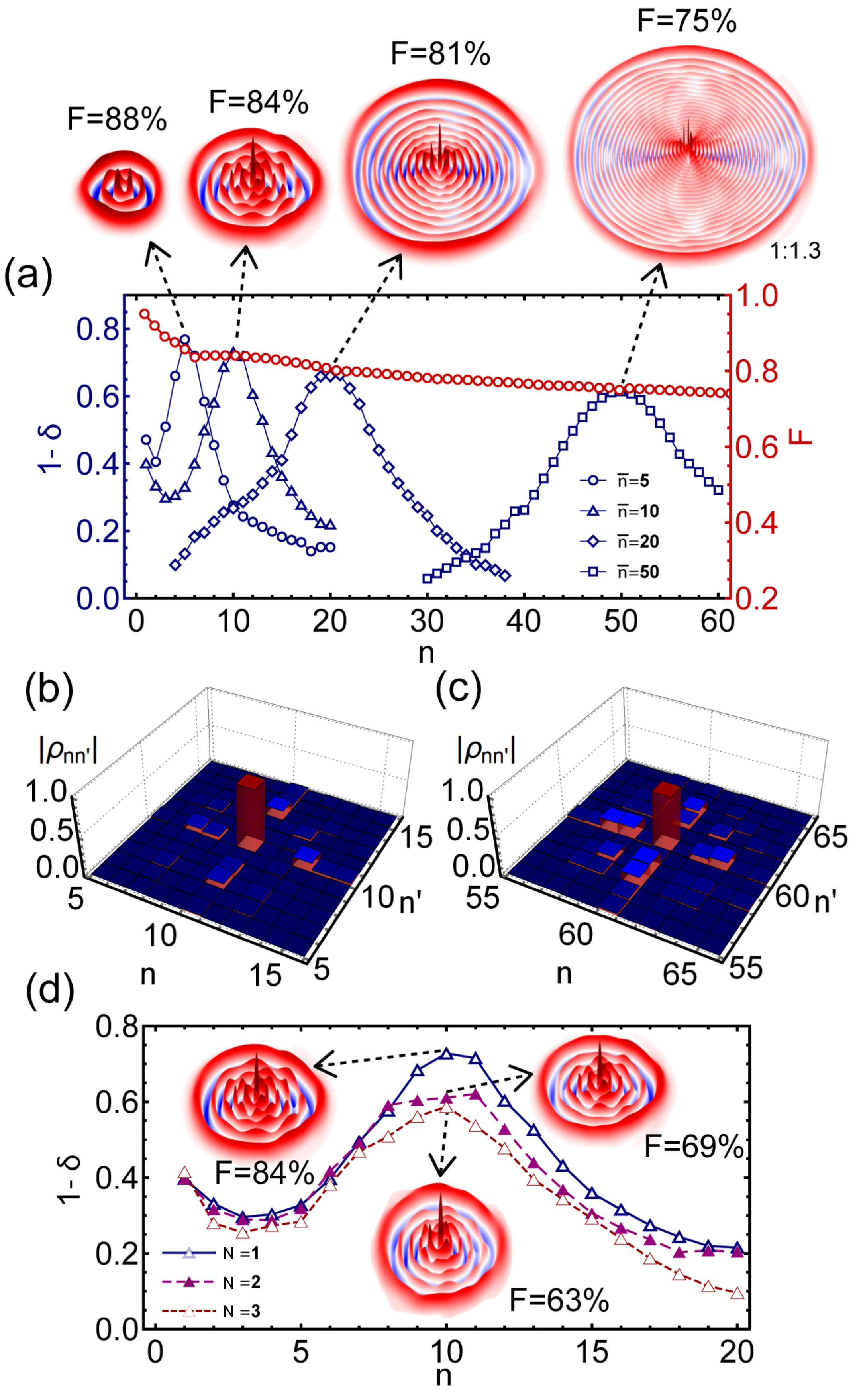}
 \caption{(a) The left blue axis shows one minus the trace distance between the obtained state and $\ket{n}$ as a function of $n$, where a single two-level atom initially in $\ket{e}$ interacts with a coherent field $\alpha$ with different initial average number of photons $|\alpha|^2=\bar{n}=\left\{5,10,20,50\right\}$. The right red axis shows the obtained fidelity for $n=\bar{n}$ as a function of $n$. (b) and (c) Density matrices of the generated Fock-like states with $n=10$ and $n=60$ respectively. (d) One minus the trace distance between the obtained state and $\ket{n}$ as a function of $n$ for a field that starts in a coherent state $|\alpha|^2=\bar{n}=10$ and interacts with $N=\left\{1,2,3\right\}$ atoms. The insets on top of every curve represents the Wigner functions and fidelity of the obtained field state at $n=\bar{n}$. }
 \label{fg:II}
\end{figure}

{\it Results.---} We first consider the case of a single atom initially in the excited state interacting with a resonant coherent  field in the absence of any decoherence mechanism. Fig. \ref{fg:II}(a) shows, on the left axis, the maximum achievable $1-\delta\left(\rho_f(t,\beta),\rho_n\right)$ between the obtained field (displaced by the proper $\beta$ in each case) and a Fock state with $n$ photons as a function $n$. The calculation is repeated for different initial coherent states  with average photon number $|\alpha|^2=\bar{n}$. We observe that the optimum generation of a Fock state happens at $n=\bar{n}$, meaning that the process benefits from keeping the same average number of excitations in the field. (An example of the energy conservation throughout the full evolution of the system is presented in the Supplemental Material \cite{SM}.) For those cases, the fidelity can be higher than $75\%$ for $n \leq 50$ as shown in the right axis of Fig. \ref{fg:II}(a). Notice that the points plotted for the fidelity (with $n=\bar{n}$) correspond to the peaks of distributions like those shown in the $1-\delta$ plots as a function of $n$ for a fix $\bar{n}$. Figs. \ref{fg:II}(b) and \ref{fg:II}(c) show the elements of the density matrix for the generated Fock-like states with $n=10$ and $n=60$ respectively. We observe a small remnant in the coherences that explains the obtained fidelities despite negligible population of adjacent number states.

Collective atomic effects increase the effective interaction strength, hence shortening the necessary interaction times to generate a target Fock state. Only a few initial atomic states lead to the formation of a Fock-like state. These are linear superpositions of the eigenstates $\ket{\phi_{\lambda_{i}}}$  of the collective atomic operator $\hat{S}^{(N)}_x=\sum_i^{N}\hat{\sigma}^{(i)}_x$, i.e. $\hat{S}^{(N)}_x\ket{\phi_{\lambda_{i}}}=\lambda_i\ket{\phi_{\lambda_{i}}}$ with eigenvalues $\lambda_i$, where $\hat{\sigma}^{(i)}_x$ is the Pauli matrix operating on the $i$-atom and $N$ is the total number of atoms. In particular, the initial atomic states that lead to a Fock-like states are those of the form $\ket{\phi_{\text{at}}}=1/\sqrt{2}\left(\ket{\phi_{\lambda_{1}}}+\ket{\phi_{\lambda_{2}}}\right)$, such that $\lambda_2=-\lambda_1$.  The formation of a Fock state speeds up by a factor of $N$ when $\lambda_1=\text{max}\{\lambda_i\}$.

Figure \ref{fg:II}(d) shows one minus the trace distance of the generated Fock state, as a function of $n$, for one, two, and three atoms. The initial atomic states are $|\phi_{\text{at}}\rangle=\left\{|e_1\rangle;~\frac{1}{\sqrt{2}}\left(|g_1g_2\rangle+|e_1e_2\rangle\right)\right.;$ $\left.\frac{1}{2}\left(|e_1e_2e_3\rangle+|e_1g_2g_3\rangle+|g_1e_2g_3\rangle+|g_1g_2e_3\rangle\right)\right\}$, respectively, where $g_i$ and $e_i$ represent the $i$th-atom being in the ground or excited state respectively. We see that increasing the number of atoms has a detrimental effect on the fidelity of the final Fock state, because the purity of the field gets compromised when a larger fraction of the total coherence of the system remains in the atomic subsystem~\cite{SM}. Considering this, and the technical difficulties of realizing entangled states of many particles, we limit our analysis to the case of one and two atoms.

\begin{figure}
 \centering
 \includegraphics[width=0.48\textwidth]{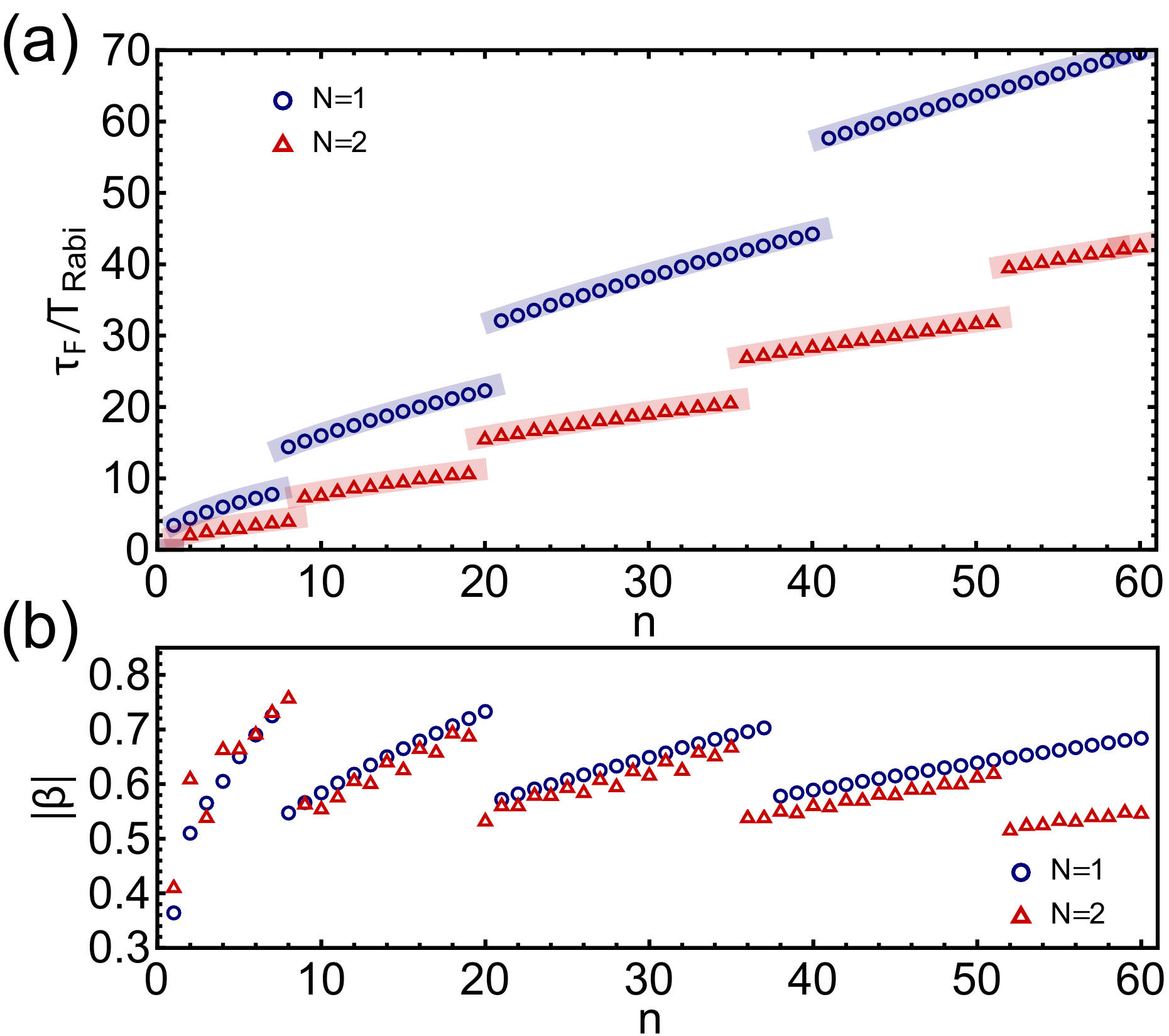}
 \caption{(a) Optimum times $\tau_{\text{F}}$ to generate a Fock state $\ket{n}$ as a function of $n$, starting from a coherent state $|\alpha|^2=\bar{n}=n$, for the case of a single atom in $|e_1\rangle$ (blue circles) and two entangled atoms in $(\vert e_1e_1\rangle + \vert g_1g_1\rangle)/\sqrt{2}$ (red triangles). The vertical axis is in units of resonant Rabi periods. (b) Displacements $\beta_\text{F}$ for the states achieved in (a) as a function of $n$.}
\label{fg:III_NEW}
\end{figure}

Figure \ref{fg:III_NEW}(a) shows the optimum times $\tau_{\text{F}}$ in units of a single-atom/single-photon Rabi period as a function of the target number of photons $n$, for one and two atoms (same initial atomic states as before). The optimal evolution time $\tau_{\text{F}}$ follows even multiples of a $(\sqrt{n}+\sqrt{n+1})$ dependence~\cite{SM}, represented by segmented shaded lines. The multiple branches appear because the Fock-like state is periodically generated throughout the evolution of the system, but with slightly different fidelities. The maximization of the fidelity leads to what looks like jumps of $\tau_{\text{F}}$ between different branches \cite{SM}. Figure \ref{fg:III_NEW}(b) shows the displacements $\beta_{\text{F}}$ necessary to generate the Fock-like state with the largest fidelity as a function of $n$. In our case the displacement is always real, since we begin the interaction with a real $\alpha$. If $\alpha$ were complex, then the appropriate displacement will have the same complex phase than $\alpha$. The role of the coherent displacement $D(\beta_{\text{F}})$ is to compensate for the energy difference between the initial/ target state and the final state of the field after energy exchange with the atoms~\cite{SM}.-  

\begin{figure}
 \centering
 \includegraphics[width=0.48\textwidth]{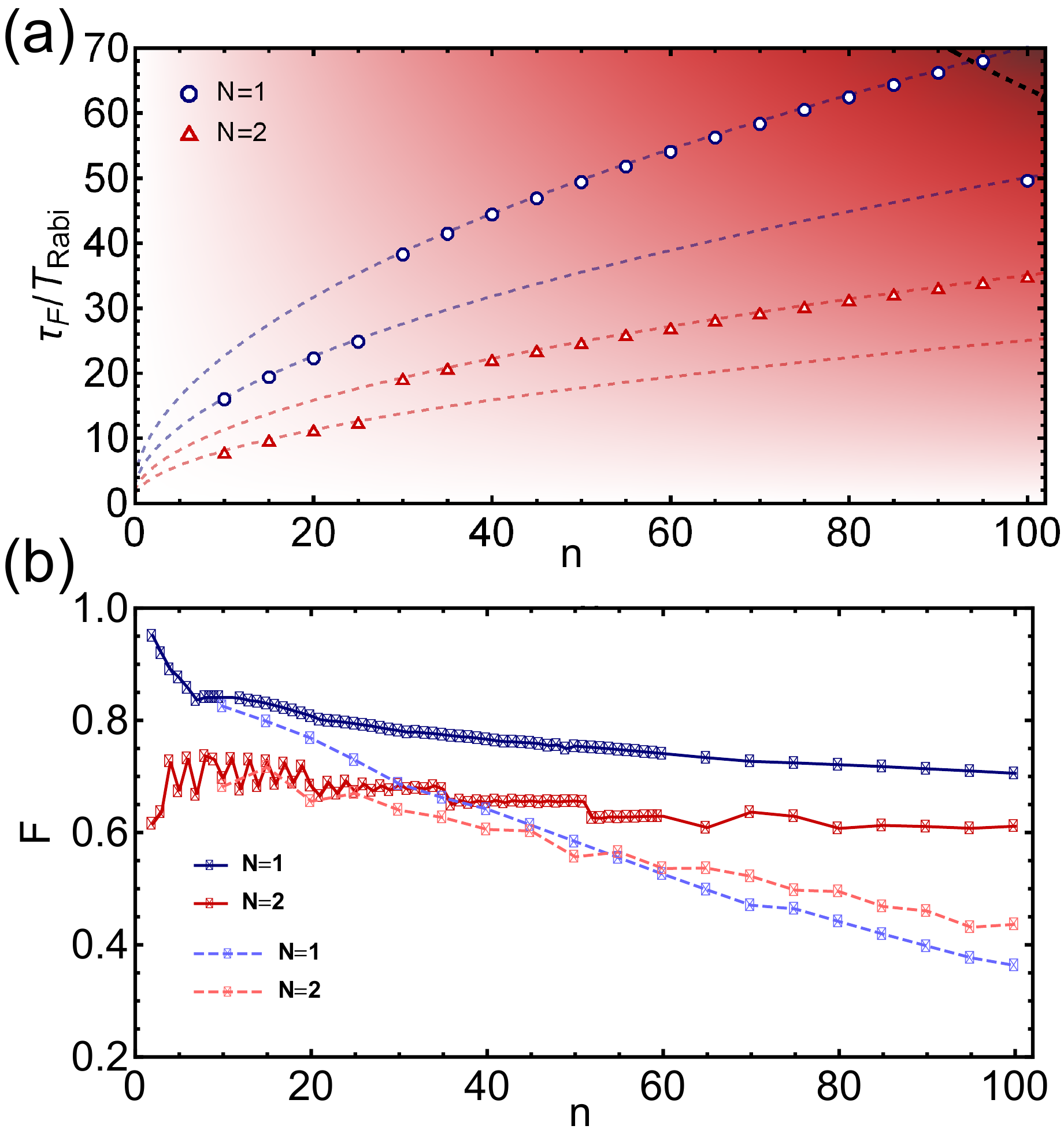}
 \caption{(a) Optimum times $\tau_{\text{F}}$ to generate a Fock state $\ket{n}$ as a function of $n$, starting from a coherent state $|\alpha|^2=\bar{n}=n$, for the one and two atom cases (see Fig. \ref{fg:II}). The vertical axis is in units of resonant Rabi periods. Colored dashed line corresponds to the multiple branches of the solutions for both cases~\cite{SM}. The black dashed line represents the decay time of a Fock state of $n$ photons inside the cavity. (b) Fidelities for the states generated in (a) as a function of $n$. Continuous lines denote the ideal lossless system, and dashed lines denotes the system under realistic decoherence mechanisms in CQED.}\label{fg:III}
\end{figure}

{\it Experimental feasibility in CQED.---} We analyse our protocol for typical experimental parameters in CQED with Rydberg atoms~\cite{zhou2012PRL} (see Fig. \ref{fg:setup}). We consider decoherence from cavity and atomic losses, and thermal photons (Eq. (\ref{eq:master})) to be $\kappa=1/T_c$, with $T_c=130$~ms the cavity damping time, $\Gamma=1/T_a$, with $T_a=30$~ms the atomic lifetime, $n_{\text{th}}=0.05$ (at $0.8$~K) \cite{guerlin2007}, and a vacuum Rabi frequency of $\Omega_0=2 \pi \times 49$~kHz \cite{HarocheBook}. The atoms are sent through the cavity as a jet, and the interaction time is controlled by the atomic velocity. 
Fig. \ref{fg:III}(a) shows the optimum times $\tau_{\text{F}}$ as a function of $n$, for one and two atoms. The decay rate of a Fock states with $n$ photons is $\kappa n$ \cite{Wang2008}, represented by the dashed line and a shaded area in Fig. \ref{fg:III}(a). The extension of the scheme to larger photon number states or larger number of atoms is truncated by decoherence effects, and its exploration is limited by our computational capabilities. Figure \ref{fg:III}(b) shows the maximum fidelity for one and two atoms as a function of the target Fock state, both in the ideal case and in the presence of decoherence. Notice that in the presence of decoherence one benefits of using two atoms for target Fock states above $n=50$. The obtained Fock-like states are robust against imperfections in both the evolution time $\tau_{\text{F}}$ and the coherent displacement $\beta_{\text{F}}$, where typical experimental errors produce negligible changes in the state fidelity \cite{SM,SayrinThesis}.

State of the art of CQED experiments with Rydberg atoms can reach a maximum interaction time of $20$ Rabi periods \cite{sebastien2019}. Although experimental improvements are being made on that regard, this presents an opportunity to study protocols for state preparation of a few entangled atoms to speed up the state generation process. 

We observe that even for the simplest of the previously described cases, meaning a single excited atom interacting with a coherent state, it is possible to achieve larger fidelities by conditioning the field to a particular post-selected atomic state \cite{HarocheBook}. As a comparison, for an initial/target state $\bar{n}=n=10$ and a single atom we obtain a fidelity of $92\%$ upon measuring the atom in the excited state, compared with $F=84\%$ in the un-projected case. A careful analysis of optimal projections on the atomic subsystem goes beyond the scope of this paper, but it opens an opportunity to improve the presented scheme.

Our protocol for Fock states generation can be easily extended to other platforms. Circuit-QED systems are particularly interesting, since the interaction time between the artificial atom and the field is arbitrarily large. Considering parameters of state of the art circuit-QED experiments \cite{Campagne2019,Kjaergaard2019,Landig2019}, it is possible to generate a Fock state near $n=100$ with $60\%$ fidelity with a single qubit, close to the performance of the protocol without decoherence.

{\it Analysis and Discussion.---} We observe that the generation of Fock-like states requires two physical phenomena: non-linearity and interference. Given that the effective Rabi frequency depends on the photon number as $\Omega=\Omega_0\sqrt{n}$, the probability distribution of the coherent state will be distorted upon unitary evolution, resulting in the negativity of its Wigner function. This is a particular case of a non-linear evolution generating a non-classical state~\cite{Albarelli2016}. On the other hand, when the interacting two-level system is in a superposition, the field evolves as such, allowing for interference effects among probability amplitudes of the field overlapping in phase space (see video in \cite{SM}). The non-linear $\hat{n}$ dependence and the ability to evolve the field in a superposition, generating Fock-like states, are not unique features of the Jaynes-Cummings model. For example, the effective interaction Hamiltonian $H_{\text{eff}}=g/2\sqrt{\hat{n}}\hat{S}_x^{(N)}$~\cite{Saavedra1997,AnaMaria2019} can also generate Fock-like states. While the non-linearity-plus-interference picture explains the generation of highly non-classical states, it does not answer why we can obtain Fock-like states in particular or why these have such large and robust fidelities. Nonetheless, it is not too surprising that the distribution of a field unitarily evolved to have a large phase uncertainty will resemble that of a Fock state. The most remarkable aspect of the presented protocol is the fact that a macroscopically intense classical field can express its truly granular (quantum) nature upon interaction with a quantum two-level system.

The presented scheme could be implemented across different QED platforms. Its limiting factors is the shortest time scale for the decoherence. Since the system needs to undergo several Rabi cycles before the fields ends in a Fock-like state, we suspect that the necessary condition to succeed is that of strong coupling, where $g\gg n\kappa,\Gamma$. 
 
{\it Conclusions.---} We have presented a protocol to deterministically generate large photon number state within QED systems. 
The 
field starts in a coherent state and evolves to a Fock-like state upon resonant interaction with 
a two-level system, without need of a post-selective procedure. The intrinsic non-linear evolution of the field plus interference effects in the photon number probability amplitudes generate a state of the field that is well described by a Fock state coherently displaced in phase space. After correcting for such displacement, we obtain a Fock-like state with optimal fidelities as large as $71\%$ for $n=100$. We show how this process can be sped up aided by a second two-level system, but compromising the fidelity of the final state. The scheme shows to be feasible for current state-of-the-art experiments. Although our analysis is mainly focused  on a CQED system, it can be extended to other QED platforms. We expect that the implementation of the presented protocol will have a significant impact on quantum metrology applications.\\

{\it Acknowledgments.---} 
We thank Vladan Vuleti\'{c} for helpful discussions at the genesis of this project and Jean-Michel Raimond and Bruno Peaudecerf for their useful comments and suggestions. Authors acknowledge J. Rojas for her help on figures design. This work was supported in part by FONDECYT Grant N$^{\circ}$ 11190078, CONICYT-PAI grants 77190033 and 77180003, and Programa ICM Millennium Institute for Research in Optics (MIRO). All numerical calculations were performed with Julia Programming Language and libraries within.

\bibliography{References}

\begin{thebibliography}{48}%
\makeatletter
\providecommand \@ifxundefined [1]{%
 \@ifx{#1\undefined}
}%
\providecommand \@ifnum [1]{%
 \ifnum #1\expandafter \@firstoftwo
 \else \expandafter \@secondoftwo
 \fi
}%
\providecommand \@ifx [1]{%
 \ifx #1\expandafter \@firstoftwo
 \else \expandafter \@secondoftwo
 \fi
}%
\providecommand \natexlab [1]{#1}%
\providecommand \enquote  [1]{``#1''}%
\providecommand \bibnamefont  [1]{#1}%
\providecommand \bibfnamefont [1]{#1}%
\providecommand \citenamefont [1]{#1}%
\providecommand \href@noop [0]{\@secondoftwo}%
\providecommand \href [0]{\begingroup \@sanitize@url \@href}%
\providecommand \@href[1]{\@@startlink{#1}\@@href}%
\providecommand \@@href[1]{\endgroup#1\@@endlink}%
\providecommand \@sanitize@url [0]{\catcode `\\12\catcode `\$12\catcode
  `\&12\catcode `\#12\catcode `\^12\catcode `\_12\catcode `\%12\relax}%
\providecommand \@@startlink[1]{}%
\providecommand \@@endlink[0]{}%
\providecommand \url  [0]{\begingroup\@sanitize@url \@url }%
\providecommand \@url [1]{\endgroup\@href {#1}{\urlprefix }}%
\providecommand \urlprefix  [0]{URL }%
\providecommand \Eprint [0]{\href }%
\providecommand \doibase [0]{http://dx.doi.org/}%
\providecommand \selectlanguage [0]{\@gobble}%
\providecommand \bibinfo  [0]{\@secondoftwo}%
\providecommand \bibfield  [0]{\@secondoftwo}%
\providecommand \translation [1]{[#1]}%
\providecommand \BibitemOpen [0]{}%
\providecommand \bibitemStop [0]{}%
\providecommand \bibitemNoStop [0]{.\EOS\space}%
\providecommand \EOS [0]{\spacefactor3000\relax}%
\providecommand \BibitemShut  [1]{\csname bibitem#1\endcsname}%
\let\auto@bib@innerbib\@empty
\bibitem [{\citenamefont {{S. Haroche and J.M. Raimond}}(2013)}]{HarocheBook}%
  \BibitemOpen
  \bibfield  {author} {\bibinfo {author} {\bibnamefont {{S. Haroche and J.M.
  Raimond}}},\ }\href@noop {} {\bibfield  {journal} {\bibinfo  {journal}
  {Oxford University Press}\ } (\bibinfo {year} {2013})}\BibitemShut {NoStop}%
\bibitem [{\citenamefont {{Giovannetti, V. and Lloyd, S. and Maccone,
  L.}}(2011)}]{Giovannetti_2011}%
  \BibitemOpen
  \bibfield  {author} {\bibinfo {author} {\bibnamefont {{Giovannetti, V. and
  Lloyd, S. and Maccone, L.}}},\ }\href {\doibase 10.1038/nphoton.2011.35}
  {\bibfield  {journal} {\bibinfo  {journal} {Nature Photonics}\ }\textbf
  {\bibinfo {volume} {5}},\ \bibinfo {pages} {222} (\bibinfo {year}
  {2011})}\BibitemShut {NoStop}%
\bibitem [{\citenamefont {Varcoe}\ \emph {et~al.}(2000)\citenamefont {Varcoe},
  \citenamefont {Brattke}, \citenamefont {Weidinger},\ and\ \citenamefont
  {Walther}}]{varcoe2000}%
  \BibitemOpen
  \bibfield  {author} {\bibinfo {author} {\bibfnamefont {B.~T.~H.}\
  \bibnamefont {Varcoe}}, \bibinfo {author} {\bibfnamefont {S.}~\bibnamefont
  {Brattke}}, \bibinfo {author} {\bibfnamefont {M.}~\bibnamefont {Weidinger}},
  \ and\ \bibinfo {author} {\bibfnamefont {H.}~\bibnamefont {Walther}},\ }\href
  {\doibase 10.1038/35001526} {\bibfield  {journal} {\bibinfo  {journal}
  {Nature}\ }\textbf {\bibinfo {volume} {403}},\ \bibinfo {pages} {743}
  (\bibinfo {year} {2000})}\BibitemShut {NoStop}%
\bibitem [{\citenamefont {Brattke}\ \emph {et~al.}(2001)\citenamefont
  {Brattke}, \citenamefont {Varcoe},\ and\ \citenamefont
  {Walther}}]{Brattke2001}%
  \BibitemOpen
  \bibfield  {author} {\bibinfo {author} {\bibfnamefont {S.}~\bibnamefont
  {Brattke}}, \bibinfo {author} {\bibfnamefont {B.~T.~H.}\ \bibnamefont
  {Varcoe}}, \ and\ \bibinfo {author} {\bibfnamefont {H.}~\bibnamefont
  {Walther}},\ }\href {\doibase 10.1103/PhysRevLett.86.3534} {\bibfield
  {journal} {\bibinfo  {journal} {Phys. Rev. Lett.}\ }\textbf {\bibinfo
  {volume} {86}},\ \bibinfo {pages} {3534} (\bibinfo {year}
  {2001})}\BibitemShut {NoStop}%
\bibitem [{\citenamefont {Bertet}\ \emph {et~al.}(2002)\citenamefont {Bertet},
  \citenamefont {Osnaghi}, \citenamefont {Milman}, \citenamefont {Auffeves},
  \citenamefont {Maioli}, \citenamefont {Brune}, \citenamefont {Raimond},\ and\
  \citenamefont {Haroche}}]{Bertet2002}%
  \BibitemOpen
  \bibfield  {author} {\bibinfo {author} {\bibfnamefont {P.}~\bibnamefont
  {Bertet}}, \bibinfo {author} {\bibfnamefont {S.}~\bibnamefont {Osnaghi}},
  \bibinfo {author} {\bibfnamefont {P.}~\bibnamefont {Milman}}, \bibinfo
  {author} {\bibfnamefont {A.}~\bibnamefont {Auffeves}}, \bibinfo {author}
  {\bibfnamefont {P.}~\bibnamefont {Maioli}}, \bibinfo {author} {\bibfnamefont
  {M.}~\bibnamefont {Brune}}, \bibinfo {author} {\bibfnamefont {J.~M.}\
  \bibnamefont {Raimond}}, \ and\ \bibinfo {author} {\bibfnamefont
  {S.}~\bibnamefont {Haroche}},\ }\href {\doibase
  10.1103/PhysRevLett.88.143601} {\bibfield  {journal} {\bibinfo  {journal}
  {Phys. Rev. Lett.}\ }\textbf {\bibinfo {volume} {88}},\ \bibinfo {pages}
  {143601} (\bibinfo {year} {2002})}\BibitemShut {NoStop}%
\bibitem [{\citenamefont {Waks}\ \emph {et~al.}(2006)\citenamefont {Waks},
  \citenamefont {Diamanti},\ and\ \citenamefont {Yamamoto}}]{Waks2006}%
  \BibitemOpen
  \bibfield  {author} {\bibinfo {author} {\bibfnamefont {E.}~\bibnamefont
  {Waks}}, \bibinfo {author} {\bibfnamefont {E.}~\bibnamefont {Diamanti}}, \
  and\ \bibinfo {author} {\bibfnamefont {Y.}~\bibnamefont {Yamamoto}},\ }\href
  {https://doi.org/10.1088/1367-2630/8/1/004} {\bibfield  {journal} {\bibinfo
  {journal} {New Journal of Physics}\ }\textbf {\bibinfo {volume} {8}}
  (\bibinfo {year} {2006})}\BibitemShut {NoStop}%
\bibitem [{\citenamefont {{Hofheinz}}\ \emph {et~al.}(2008)\citenamefont
  {{Hofheinz}}, \citenamefont {{Weig}}, \citenamefont {{Ansmann}},
  \citenamefont {{Bialczak}}, \citenamefont {{Lucero}}, \citenamefont
  {{Neeley}}, \citenamefont {{O'Connell}}, \citenamefont {{Wang}},
  \citenamefont {{Martinis}},\ and\ \citenamefont
  {{Cleland}}}]{2008Natur.454..310H}%
  \BibitemOpen
  \bibfield  {author} {\bibinfo {author} {\bibfnamefont {M.}~\bibnamefont
  {{Hofheinz}}}, \bibinfo {author} {\bibfnamefont {E.~M.}\ \bibnamefont
  {{Weig}}}, \bibinfo {author} {\bibfnamefont {M.}~\bibnamefont {{Ansmann}}},
  \bibinfo {author} {\bibfnamefont {R.~C.}\ \bibnamefont {{Bialczak}}},
  \bibinfo {author} {\bibfnamefont {E.}~\bibnamefont {{Lucero}}}, \bibinfo
  {author} {\bibfnamefont {M.}~\bibnamefont {{Neeley}}}, \bibinfo {author}
  {\bibfnamefont {A.~D.}\ \bibnamefont {{O'Connell}}}, \bibinfo {author}
  {\bibfnamefont {H.}~\bibnamefont {{Wang}}}, \bibinfo {author} {\bibfnamefont
  {J.~M.}\ \bibnamefont {{Martinis}}}, \ and\ \bibinfo {author} {\bibfnamefont
  {A.~N.}\ \bibnamefont {{Cleland}}},\ }\href {\doibase 10.1038/nature07136}
  {\bibfield  {journal} {\bibinfo  {journal} {\nat}\ }\textbf {\bibinfo
  {volume} {454}},\ \bibinfo {pages} {310} (\bibinfo {year}
  {2008})}\BibitemShut {NoStop}%
\bibitem [{\citenamefont {Chu}\ \emph {et~al.}(2018)\citenamefont {Chu},
  \citenamefont {Kharel}, \citenamefont {Yoon}, \citenamefont {Frunzio},
  \citenamefont {Rakich},\ and\ \citenamefont {Schoelkopf}}]{Chu2018}%
  \BibitemOpen
  \bibfield  {author} {\bibinfo {author} {\bibfnamefont {Y.}~\bibnamefont
  {Chu}}, \bibinfo {author} {\bibfnamefont {P.}~\bibnamefont {Kharel}},
  \bibinfo {author} {\bibfnamefont {T.}~\bibnamefont {Yoon}}, \bibinfo {author}
  {\bibfnamefont {L.}~\bibnamefont {Frunzio}}, \bibinfo {author} {\bibfnamefont
  {P.~T.}\ \bibnamefont {Rakich}}, \ and\ \bibinfo {author} {\bibfnamefont
  {R.~J.}\ \bibnamefont {Schoelkopf}},\ }\href {\doibase
  10.1038/s41586-018-0717-7} {\bibfield  {journal} {\bibinfo  {journal}
  {Nature}\ }\textbf {\bibinfo {volume} {563}},\ \bibinfo {pages} {666}
  (\bibinfo {year} {2018})}\BibitemShut {NoStop}%
\bibitem [{\citenamefont {Tiedau}\ \emph {et~al.}(2019)\citenamefont {Tiedau},
  \citenamefont {Bartley}, \citenamefont {Harder}, \citenamefont {Lita},
  \citenamefont {Nam}, \citenamefont {Gerrits},\ and\ \citenamefont
  {Silberhorn}}]{Tiedau2019}%
  \BibitemOpen
  \bibfield  {author} {\bibinfo {author} {\bibfnamefont {J.}~\bibnamefont
  {Tiedau}}, \bibinfo {author} {\bibfnamefont {T.~J.}\ \bibnamefont {Bartley}},
  \bibinfo {author} {\bibfnamefont {G.}~\bibnamefont {Harder}}, \bibinfo
  {author} {\bibfnamefont {A.~E.}\ \bibnamefont {Lita}}, \bibinfo {author}
  {\bibfnamefont {S.~W.}\ \bibnamefont {Nam}}, \bibinfo {author} {\bibfnamefont
  {T.}~\bibnamefont {Gerrits}}, \ and\ \bibinfo {author} {\bibfnamefont
  {C.}~\bibnamefont {Silberhorn}},\ }\href {\doibase
  10.1103/PhysRevA.100.041802} {\bibfield  {journal} {\bibinfo  {journal}
  {Phys. Rev. A}\ }\textbf {\bibinfo {volume} {100}},\ \bibinfo {pages}
  {041802} (\bibinfo {year} {2019})}\BibitemShut {NoStop}%
\bibitem [{\citenamefont {Yanagimoto}\ \emph {et~al.}(2019)\citenamefont
  {Yanagimoto}, \citenamefont {Ng}, \citenamefont {Onodera},\ and\
  \citenamefont {Mabuchi}}]{Yanagimoto2019}%
  \BibitemOpen
  \bibfield  {author} {\bibinfo {author} {\bibfnamefont {R.}~\bibnamefont
  {Yanagimoto}}, \bibinfo {author} {\bibfnamefont {E.}~\bibnamefont {Ng}},
  \bibinfo {author} {\bibfnamefont {T.}~\bibnamefont {Onodera}}, \ and\
  \bibinfo {author} {\bibfnamefont {H.}~\bibnamefont {Mabuchi}},\ }\href
  {\doibase 10.1103/PhysRevA.100.033822} {\bibfield  {journal} {\bibinfo
  {journal} {Phys. Rev. A}\ }\textbf {\bibinfo {volume} {100}},\ \bibinfo
  {pages} {033822} (\bibinfo {year} {2019})}\BibitemShut {NoStop}%
\bibitem [{\citenamefont {Dell’Anno}\ \emph {et~al.}(2006)\citenamefont
  {Dell’Anno}, \citenamefont {Silvio}, \citenamefont {Siena},\ and\
  \citenamefont {Illuminati}}]{DellAnno2006}%
  \BibitemOpen
  \bibfield  {author} {\bibinfo {author} {\bibfnamefont {F.}~\bibnamefont
  {Dell’Anno}}, \bibinfo {author} {\bibnamefont {Silvio}}, \bibinfo {author}
  {\bibfnamefont {D.}~\bibnamefont {Siena}}, \ and\ \bibinfo {author}
  {\bibfnamefont {F.}~\bibnamefont {Illuminati}},\ }\href
  {https://doi.org/10.1016/j.physrep.2006.01.004} {\bibfield  {journal}
  {\bibinfo  {journal} {Physics Reports}\ }\textbf {\bibinfo {volume} {53}}
  (\bibinfo {year} {2006})}\BibitemShut {NoStop}%
\bibitem [{\citenamefont {Gonz\'alez-Tudela}\ \emph {et~al.}(2015)\citenamefont
  {Gonz\'alez-Tudela}, \citenamefont {Paulisch}, \citenamefont {Chang},
  \citenamefont {Kimble},\ and\ \citenamefont {Cirac}}]{GonzalesTudela2015}%
  \BibitemOpen
  \bibfield  {author} {\bibinfo {author} {\bibfnamefont {A.}~\bibnamefont
  {Gonz\'alez-Tudela}}, \bibinfo {author} {\bibfnamefont {V.}~\bibnamefont
  {Paulisch}}, \bibinfo {author} {\bibfnamefont {D.~E.}\ \bibnamefont {Chang}},
  \bibinfo {author} {\bibfnamefont {H.~J.}\ \bibnamefont {Kimble}}, \ and\
  \bibinfo {author} {\bibfnamefont {J.~I.}\ \bibnamefont {Cirac}},\ }\href
  {\doibase 10.1103/PhysRevLett.115.163603} {\bibfield  {journal} {\bibinfo
  {journal} {Phys. Rev. Lett.}\ }\textbf {\bibinfo {volume} {115}},\ \bibinfo
  {pages} {163603} (\bibinfo {year} {2015})}\BibitemShut {NoStop}%
\bibitem [{\citenamefont {Gonz\'alez-Tudela}\ \emph {et~al.}(2017)\citenamefont
  {Gonz\'alez-Tudela}, \citenamefont {Paulisch}, \citenamefont {Kimble},\ and\
  \citenamefont {Cirac}}]{GonzalesTudela2017}%
  \BibitemOpen
  \bibfield  {author} {\bibinfo {author} {\bibfnamefont {A.}~\bibnamefont
  {Gonz\'alez-Tudela}}, \bibinfo {author} {\bibfnamefont {V.}~\bibnamefont
  {Paulisch}}, \bibinfo {author} {\bibfnamefont {H.~J.}\ \bibnamefont
  {Kimble}}, \ and\ \bibinfo {author} {\bibfnamefont {J.~I.}\ \bibnamefont
  {Cirac}},\ }\href {\doibase 10.1103/PhysRevLett.118.213601} {\bibfield
  {journal} {\bibinfo  {journal} {Phys. Rev. Lett.}\ }\textbf {\bibinfo
  {volume} {118}},\ \bibinfo {pages} {213601} (\bibinfo {year}
  {2017})}\BibitemShut {NoStop}%
\bibitem [{\citenamefont {Perarnau-Llobet}\ \emph {et~al.}(2020)\citenamefont
  {Perarnau-Llobet}, \citenamefont {Gonz{\'{a}}lez-Tudela},\ and\ \citenamefont
  {Cirac}}]{GonzalesTudela2020}%
  \BibitemOpen
  \bibfield  {author} {\bibinfo {author} {\bibfnamefont {M.}~\bibnamefont
  {Perarnau-Llobet}}, \bibinfo {author} {\bibfnamefont {A.}~\bibnamefont
  {Gonz{\'{a}}lez-Tudela}}, \ and\ \bibinfo {author} {\bibfnamefont {J.~I.}\
  \bibnamefont {Cirac}},\ }\href {\doibase 10.1088/2058-9565/ab6ce5} {\bibfield
   {journal} {\bibinfo  {journal} {Quantum Science and Technology}\ }\textbf
  {\bibinfo {volume} {5}},\ \bibinfo {pages} {025003} (\bibinfo {year}
  {2020})}\BibitemShut {NoStop}%
\bibitem [{\citenamefont {Leek}\ \emph {et~al.}(2010)\citenamefont {Leek},
  \citenamefont {Baur}, \citenamefont {Fink}, \citenamefont {Bianchetti},
  \citenamefont {Steffen}, \citenamefont {Filipp},\ and\ \citenamefont
  {Wallraff}}]{Leek2010}%
  \BibitemOpen
  \bibfield  {author} {\bibinfo {author} {\bibfnamefont {P.~J.}\ \bibnamefont
  {Leek}}, \bibinfo {author} {\bibfnamefont {M.}~\bibnamefont {Baur}}, \bibinfo
  {author} {\bibfnamefont {J.~M.}\ \bibnamefont {Fink}}, \bibinfo {author}
  {\bibfnamefont {R.}~\bibnamefont {Bianchetti}}, \bibinfo {author}
  {\bibfnamefont {L.}~\bibnamefont {Steffen}}, \bibinfo {author} {\bibfnamefont
  {S.}~\bibnamefont {Filipp}}, \ and\ \bibinfo {author} {\bibfnamefont
  {A.}~\bibnamefont {Wallraff}},\ }\href {\doibase
  10.1103/PhysRevLett.104.100504} {\bibfield  {journal} {\bibinfo  {journal}
  {Phys. Rev. Lett.}\ }\textbf {\bibinfo {volume} {104}},\ \bibinfo {pages}
  {100504} (\bibinfo {year} {2010})}\BibitemShut {NoStop}%
\bibitem [{\citenamefont {Premaratne}\ \emph {et~al.}(2017)\citenamefont
  {Premaratne}, \citenamefont {Wellstood},\ and\ \citenamefont
  {Palmer}}]{Premaratne2017}%
  \BibitemOpen
  \bibfield  {author} {\bibinfo {author} {\bibfnamefont {S.~P.}\ \bibnamefont
  {Premaratne}}, \bibinfo {author} {\bibfnamefont {F.~C.}\ \bibnamefont
  {Wellstood}}, \ and\ \bibinfo {author} {\bibfnamefont {B.~S.}\ \bibnamefont
  {Palmer}},\ }\href {\doibase 10.1038/ncomms14148} {\bibfield  {journal}
  {\bibinfo  {journal} {Nature Communications}\ }\textbf {\bibinfo {volume}
  {8}},\ \bibinfo {pages} {14148} (\bibinfo {year} {2017})}\BibitemShut
  {NoStop}%
\bibitem [{\citenamefont {Gu}\ \emph {et~al.}(2017)\citenamefont {Gu},
  \citenamefont {Kockum}, \citenamefont {Miranowicz}, \citenamefont {Liu},\
  and\ \citenamefont {Nori}}]{gu2017}%
  \BibitemOpen
  \bibfield  {author} {\bibinfo {author} {\bibfnamefont {X.}~\bibnamefont
  {Gu}}, \bibinfo {author} {\bibfnamefont {A.~F.}\ \bibnamefont {Kockum}},
  \bibinfo {author} {\bibfnamefont {A.}~\bibnamefont {Miranowicz}}, \bibinfo
  {author} {\bibfnamefont {Y.-x.}\ \bibnamefont {Liu}}, \ and\ \bibinfo
  {author} {\bibfnamefont {F.}~\bibnamefont {Nori}},\ }\href {\doibase
  10.1016/j.physrep.2017.10.002} {\bibfield  {journal} {\bibinfo  {journal}
  {Physics Reports}\ }\textbf {\bibinfo {volume} {718}},\ \bibinfo {pages} {1}
  (\bibinfo {year} {2017})}\BibitemShut {NoStop}%
\bibitem [{\citenamefont {Hofheinz}\ \emph {et~al.}(2009)\citenamefont
  {Hofheinz}, \citenamefont {Wang}, \citenamefont {Ansmann}, \citenamefont
  {Bialczak}, \citenamefont {Lucero}, \citenamefont {Neeley}, \citenamefont
  {O'connell}, \citenamefont {Sank}, \citenamefont {Wenner}, \citenamefont
  {Martinis} \emph {et~al.}}]{hofheinz2009}%
  \BibitemOpen
  \bibfield  {author} {\bibinfo {author} {\bibfnamefont {M.}~\bibnamefont
  {Hofheinz}}, \bibinfo {author} {\bibfnamefont {H.}~\bibnamefont {Wang}},
  \bibinfo {author} {\bibfnamefont {M.}~\bibnamefont {Ansmann}}, \bibinfo
  {author} {\bibfnamefont {R.~C.}\ \bibnamefont {Bialczak}}, \bibinfo {author}
  {\bibfnamefont {E.}~\bibnamefont {Lucero}}, \bibinfo {author} {\bibfnamefont
  {M.}~\bibnamefont {Neeley}}, \bibinfo {author} {\bibfnamefont
  {A.}~\bibnamefont {O'connell}}, \bibinfo {author} {\bibfnamefont
  {D.}~\bibnamefont {Sank}}, \bibinfo {author} {\bibfnamefont {J.}~\bibnamefont
  {Wenner}}, \bibinfo {author} {\bibfnamefont {J.~M.}\ \bibnamefont
  {Martinis}},  \emph {et~al.},\ }\href {\doibase 10.1038/nature08005}
  {\bibfield  {journal} {\bibinfo  {journal} {Nature}\ }\textbf {\bibinfo
  {volume} {459}},\ \bibinfo {pages} {546} (\bibinfo {year}
  {2009})}\BibitemShut {NoStop}%
\bibitem [{\citenamefont {Vogel}\ \emph {et~al.}(1993)\citenamefont {Vogel},
  \citenamefont {Akulin},\ and\ \citenamefont {Schleich}}]{Vogel1993}%
  \BibitemOpen
  \bibfield  {author} {\bibinfo {author} {\bibfnamefont {K.}~\bibnamefont
  {Vogel}}, \bibinfo {author} {\bibfnamefont {V.~M.}\ \bibnamefont {Akulin}}, \
  and\ \bibinfo {author} {\bibfnamefont {W.~P.}\ \bibnamefont {Schleich}},\
  }\href {\doibase 10.1103/PhysRevLett.71.1816} {\bibfield  {journal} {\bibinfo
   {journal} {Phys. Rev. Lett.}\ }\textbf {\bibinfo {volume} {71}},\ \bibinfo
  {pages} {1816} (\bibinfo {year} {1993})}\BibitemShut {NoStop}%
\bibitem [{\citenamefont {Meystre}\ \emph {et~al.}(1988)\citenamefont
  {Meystre}, \citenamefont {Rempe},\ and\ \citenamefont
  {Walther}}]{Meystre:88}%
  \BibitemOpen
  \bibfield  {author} {\bibinfo {author} {\bibfnamefont {P.}~\bibnamefont
  {Meystre}}, \bibinfo {author} {\bibfnamefont {G.}~\bibnamefont {Rempe}}, \
  and\ \bibinfo {author} {\bibfnamefont {H.}~\bibnamefont {Walther}},\ }\href
  {\doibase 10.1364/OL.13.001078} {\bibfield  {journal} {\bibinfo  {journal}
  {Opt. Lett.}\ }\textbf {\bibinfo {volume} {13}},\ \bibinfo {pages} {1078}
  (\bibinfo {year} {1988})}\BibitemShut {NoStop}%
\bibitem [{\citenamefont {Weidinger}\ \emph {et~al.}(1999)\citenamefont
  {Weidinger}, \citenamefont {Varcoe}, \citenamefont {Heerlein},\ and\
  \citenamefont {Walther}}]{PhysRevLett.82.3795}%
  \BibitemOpen
  \bibfield  {author} {\bibinfo {author} {\bibfnamefont {M.}~\bibnamefont
  {Weidinger}}, \bibinfo {author} {\bibfnamefont {B.~T.~H.}\ \bibnamefont
  {Varcoe}}, \bibinfo {author} {\bibfnamefont {R.}~\bibnamefont {Heerlein}}, \
  and\ \bibinfo {author} {\bibfnamefont {H.}~\bibnamefont {Walther}},\ }\href
  {\doibase 10.1103/PhysRevLett.82.3795} {\bibfield  {journal} {\bibinfo
  {journal} {Phys. Rev. Lett.}\ }\textbf {\bibinfo {volume} {82}},\ \bibinfo
  {pages} {3795} (\bibinfo {year} {1999})}\BibitemShut {NoStop}%
\bibitem [{\citenamefont {{Guerlin}}\ \emph {et~al.}(2007)\citenamefont
  {{Guerlin}}, \citenamefont {{Bernu}}, \citenamefont {{Del{\'e}glise}},
  \citenamefont {{Sayrin}}, \citenamefont {{Gleyzes}}, \citenamefont {{Kuhr}},
  \citenamefont {{Brune}}, \citenamefont {{Raimond}},\ and\ \citenamefont
  {{Haroche}}}]{2007Natur.448..889G}%
  \BibitemOpen
  \bibfield  {author} {\bibinfo {author} {\bibfnamefont {C.}~\bibnamefont
  {{Guerlin}}}, \bibinfo {author} {\bibfnamefont {J.}~\bibnamefont {{Bernu}}},
  \bibinfo {author} {\bibfnamefont {S.}~\bibnamefont {{Del{\'e}glise}}},
  \bibinfo {author} {\bibfnamefont {C.}~\bibnamefont {{Sayrin}}}, \bibinfo
  {author} {\bibfnamefont {S.}~\bibnamefont {{Gleyzes}}}, \bibinfo {author}
  {\bibfnamefont {S.}~\bibnamefont {{Kuhr}}}, \bibinfo {author} {\bibfnamefont
  {M.}~\bibnamefont {{Brune}}}, \bibinfo {author} {\bibfnamefont {J.-M.}\
  \bibnamefont {{Raimond}}}, \ and\ \bibinfo {author} {\bibfnamefont
  {S.}~\bibnamefont {{Haroche}}},\ }\href {\doibase 10.1038/nature06057}
  {\bibfield  {journal} {\bibinfo  {journal} {\nat}\ }\textbf {\bibinfo
  {volume} {448}},\ \bibinfo {pages} {889} (\bibinfo {year}
  {2007})}\BibitemShut {NoStop}%
\bibitem [{\citenamefont {{Del{\'e}glise}}\ \emph {et~al.}(2008)\citenamefont
  {{Del{\'e}glise}}, \citenamefont {{Dotsenko}}, \citenamefont {{Sayrin}},
  \citenamefont {{Bernu}}, \citenamefont {{Brune}}, \citenamefont {{Raimond}},\
  and\ \citenamefont {{Haroche}}}]{Samuel2008Nature}%
  \BibitemOpen
  \bibfield  {author} {\bibinfo {author} {\bibfnamefont {S.}~\bibnamefont
  {{Del{\'e}glise}}}, \bibinfo {author} {\bibfnamefont {I.}~\bibnamefont
  {{Dotsenko}}}, \bibinfo {author} {\bibfnamefont {C.}~\bibnamefont
  {{Sayrin}}}, \bibinfo {author} {\bibfnamefont {J.}~\bibnamefont {{Bernu}}},
  \bibinfo {author} {\bibfnamefont {M.}~\bibnamefont {{Brune}}}, \bibinfo
  {author} {\bibfnamefont {J.-M.}\ \bibnamefont {{Raimond}}}, \ and\ \bibinfo
  {author} {\bibfnamefont {S.}~\bibnamefont {{Haroche}}},\ }\href {\doibase
  10.1038/nature07288} {\bibfield  {journal} {\bibinfo  {journal} {\nat}\
  }\textbf {\bibinfo {volume} {455}},\ \bibinfo {pages} {510} (\bibinfo {year}
  {2008})}\BibitemShut {NoStop}%
\bibitem [{\citenamefont {{Sayrin}}\ \emph {et~al.}(2011)\citenamefont
  {{Sayrin}}, \citenamefont {{Dotsenko}}, \citenamefont {{Zhou}}, \citenamefont
  {{Peaudecerf}}, \citenamefont {{Rybarczyk}}, \citenamefont {{Gleyzes}},
  \citenamefont {{Rouchon}}, \citenamefont {{Mirrahimi}}, \citenamefont
  {{Amini}}, \citenamefont {{Brune}}, \citenamefont {{Raimond}},\ and\
  \citenamefont {{Haroche}}}]{2011Natur.477...73S}%
  \BibitemOpen
  \bibfield  {author} {\bibinfo {author} {\bibfnamefont {C.}~\bibnamefont
  {{Sayrin}}}, \bibinfo {author} {\bibfnamefont {I.}~\bibnamefont
  {{Dotsenko}}}, \bibinfo {author} {\bibfnamefont {X.}~\bibnamefont {{Zhou}}},
  \bibinfo {author} {\bibfnamefont {B.}~\bibnamefont {{Peaudecerf}}}, \bibinfo
  {author} {\bibfnamefont {T.}~\bibnamefont {{Rybarczyk}}}, \bibinfo {author}
  {\bibfnamefont {S.}~\bibnamefont {{Gleyzes}}}, \bibinfo {author}
  {\bibfnamefont {P.}~\bibnamefont {{Rouchon}}}, \bibinfo {author}
  {\bibfnamefont {M.}~\bibnamefont {{Mirrahimi}}}, \bibinfo {author}
  {\bibfnamefont {H.}~\bibnamefont {{Amini}}}, \bibinfo {author} {\bibfnamefont
  {M.}~\bibnamefont {{Brune}}}, \bibinfo {author} {\bibfnamefont {J.-M.}\
  \bibnamefont {{Raimond}}}, \ and\ \bibinfo {author} {\bibfnamefont
  {S.}~\bibnamefont {{Haroche}}},\ }\href {\doibase 10.1038/nature10376}
  {\bibfield  {journal} {\bibinfo  {journal} {\nat}\ }\textbf {\bibinfo
  {volume} {477}},\ \bibinfo {pages} {73} (\bibinfo {year} {2011})}\BibitemShut
  {NoStop}%
\bibitem [{\citenamefont {Zhou}\ \emph {et~al.}(2012)\citenamefont {Zhou},
  \citenamefont {Dotsenko}, \citenamefont {Peaudecerf}, \citenamefont
  {Rybarczyk}, \citenamefont {Sayrin}, \citenamefont {Gleyzes}, \citenamefont
  {Raimond}, \citenamefont {Brune},\ and\ \citenamefont
  {Haroche}}]{zhou2012PRL}%
  \BibitemOpen
  \bibfield  {author} {\bibinfo {author} {\bibfnamefont {X.}~\bibnamefont
  {Zhou}}, \bibinfo {author} {\bibfnamefont {I.}~\bibnamefont {Dotsenko}},
  \bibinfo {author} {\bibfnamefont {B.}~\bibnamefont {Peaudecerf}}, \bibinfo
  {author} {\bibfnamefont {T.}~\bibnamefont {Rybarczyk}}, \bibinfo {author}
  {\bibfnamefont {C.}~\bibnamefont {Sayrin}}, \bibinfo {author} {\bibfnamefont
  {S.}~\bibnamefont {Gleyzes}}, \bibinfo {author} {\bibfnamefont {J.~M.}\
  \bibnamefont {Raimond}}, \bibinfo {author} {\bibfnamefont {M.}~\bibnamefont
  {Brune}}, \ and\ \bibinfo {author} {\bibfnamefont {S.}~\bibnamefont
  {Haroche}},\ }\href {\doibase 10.1103/PhysRevLett.108.243602} {\bibfield
  {journal} {\bibinfo  {journal} {Phys. Rev. Lett.}\ }\textbf {\bibinfo
  {volume} {108}},\ \bibinfo {pages} {243602} (\bibinfo {year}
  {2012})}\BibitemShut {NoStop}%
\bibitem [{\citenamefont {Geremia}(2006)}]{Geremia2006}%
  \BibitemOpen
  \bibfield  {author} {\bibinfo {author} {\bibfnamefont {J.}~\bibnamefont
  {Geremia}},\ }\href {\doibase 10.1103/PhysRevLett.97.073601} {\bibfield
  {journal} {\bibinfo  {journal} {Phys. Rev. Lett.}\ }\textbf {\bibinfo
  {volume} {97}},\ \bibinfo {pages} {073601} (\bibinfo {year}
  {2006})}\BibitemShut {NoStop}%
\bibitem [{\citenamefont {Wang}\ \emph {et~al.}(2008)\citenamefont {Wang},
  \citenamefont {Hofheinz}, \citenamefont {Ansmann}, \citenamefont {Bialczak},
  \citenamefont {Lucero}, \citenamefont {Neeley}, \citenamefont {O'Connell},
  \citenamefont {Sank}, \citenamefont {Wenner}, \citenamefont {Cleland},\ and\
  \citenamefont {Martinis}}]{Wang2008}%
  \BibitemOpen
  \bibfield  {author} {\bibinfo {author} {\bibfnamefont {H.}~\bibnamefont
  {Wang}}, \bibinfo {author} {\bibfnamefont {M.}~\bibnamefont {Hofheinz}},
  \bibinfo {author} {\bibfnamefont {M.}~\bibnamefont {Ansmann}}, \bibinfo
  {author} {\bibfnamefont {R.~C.}\ \bibnamefont {Bialczak}}, \bibinfo {author}
  {\bibfnamefont {E.}~\bibnamefont {Lucero}}, \bibinfo {author} {\bibfnamefont
  {M.}~\bibnamefont {Neeley}}, \bibinfo {author} {\bibfnamefont {A.~D.}\
  \bibnamefont {O'Connell}}, \bibinfo {author} {\bibfnamefont {D.}~\bibnamefont
  {Sank}}, \bibinfo {author} {\bibfnamefont {J.}~\bibnamefont {Wenner}},
  \bibinfo {author} {\bibfnamefont {A.~N.}\ \bibnamefont {Cleland}}, \ and\
  \bibinfo {author} {\bibfnamefont {J.~M.}\ \bibnamefont {Martinis}},\ }\href
  {\doibase 10.1103/PhysRevLett.101.240401} {\bibfield  {journal} {\bibinfo
  {journal} {Phys. Rev. Lett.}\ }\textbf {\bibinfo {volume} {101}},\ \bibinfo
  {pages} {240401} (\bibinfo {year} {2008})}\BibitemShut {NoStop}%
\bibitem [{\citenamefont {Hofheinz}\ \emph {et~al.}(2008)\citenamefont
  {Hofheinz}, \citenamefont {Weig}, \citenamefont {Ansmann}, \citenamefont
  {Bialczak}, \citenamefont {Lucero}, \citenamefont {Neeley}, \citenamefont
  {O’connell}, \citenamefont {Wang}, \citenamefont {Martinis},\ and\
  \citenamefont {Cleland}}]{hofheinz2008}%
  \BibitemOpen
  \bibfield  {author} {\bibinfo {author} {\bibfnamefont {M.}~\bibnamefont
  {Hofheinz}}, \bibinfo {author} {\bibfnamefont {E.}~\bibnamefont {Weig}},
  \bibinfo {author} {\bibfnamefont {M.}~\bibnamefont {Ansmann}}, \bibinfo
  {author} {\bibfnamefont {R.~C.}\ \bibnamefont {Bialczak}}, \bibinfo {author}
  {\bibfnamefont {E.}~\bibnamefont {Lucero}}, \bibinfo {author} {\bibfnamefont
  {M.}~\bibnamefont {Neeley}}, \bibinfo {author} {\bibfnamefont
  {A.}~\bibnamefont {O’connell}}, \bibinfo {author} {\bibfnamefont
  {H.}~\bibnamefont {Wang}}, \bibinfo {author} {\bibfnamefont {J.~M.}\
  \bibnamefont {Martinis}}, \ and\ \bibinfo {author} {\bibfnamefont
  {A.}~\bibnamefont {Cleland}},\ }\href {\doibase 10.1038/nature07136}
  {\bibfield  {journal} {\bibinfo  {journal} {Nature}\ }\textbf {\bibinfo
  {volume} {454}},\ \bibinfo {pages} {310} (\bibinfo {year}
  {2008})}\BibitemShut {NoStop}%
\bibitem [{\citenamefont {Haroche}(2013)}]{Haroche2013Rev}%
  \BibitemOpen
  \bibfield  {author} {\bibinfo {author} {\bibfnamefont {S.}~\bibnamefont
  {Haroche}},\ }\href {\doibase 10.1103/RevModPhys.85.1083} {\bibfield
  {journal} {\bibinfo  {journal} {Rev. Mod. Phys.}\ }\textbf {\bibinfo {volume}
  {85}},\ \bibinfo {pages} {1083} (\bibinfo {year} {2013})}\BibitemShut
  {NoStop}%
\bibitem [{\citenamefont {Campagne-Ibarcq}\ \emph {et~al.}(2019)\citenamefont
  {Campagne-Ibarcq}, \citenamefont {Eickbusch}, \citenamefont {Touzard},
  \citenamefont {Zalys-Geller}, \citenamefont {Frattini}, \citenamefont
  {Sivak}, \citenamefont {Reinhold}, \citenamefont {Puri}, \citenamefont
  {Shankar}, \citenamefont {Schoelkopf}, \citenamefont {Frunzio}, \citenamefont
  {Mirrahimi},\ and\ \citenamefont {Devoret}}]{Campagne2019}%
  \BibitemOpen
  \bibfield  {author} {\bibinfo {author} {\bibfnamefont {P.}~\bibnamefont
  {Campagne-Ibarcq}}, \bibinfo {author} {\bibfnamefont {A.}~\bibnamefont
  {Eickbusch}}, \bibinfo {author} {\bibfnamefont {S.}~\bibnamefont {Touzard}},
  \bibinfo {author} {\bibfnamefont {E.}~\bibnamefont {Zalys-Geller}}, \bibinfo
  {author} {\bibfnamefont {N.}~\bibnamefont {Frattini}}, \bibinfo {author}
  {\bibfnamefont {V.}~\bibnamefont {Sivak}}, \bibinfo {author} {\bibfnamefont
  {P.}~\bibnamefont {Reinhold}}, \bibinfo {author} {\bibfnamefont
  {S.}~\bibnamefont {Puri}}, \bibinfo {author} {\bibfnamefont {S.}~\bibnamefont
  {Shankar}}, \bibinfo {author} {\bibfnamefont {R.}~\bibnamefont {Schoelkopf}},
  \bibinfo {author} {\bibfnamefont {L.}~\bibnamefont {Frunzio}}, \bibinfo
  {author} {\bibfnamefont {M.}~\bibnamefont {Mirrahimi}}, \ and\ \bibinfo
  {author} {\bibfnamefont {M.}~\bibnamefont {Devoret}},\ }\href
  {https://arxiv.org/abs/1907.12487} {\  (\bibinfo {year} {2019})},\ \Eprint
  {http://arxiv.org/abs/arXiv:1907.12487} {arXiv:1907.12487} \BibitemShut
  {NoStop}%
\bibitem [{\citenamefont {Kjaergaard}\ \emph {et~al.}(2020)\citenamefont
  {Kjaergaard}, \citenamefont {Schwartz}, \citenamefont {BraumÃŒller},
  \citenamefont {Krantz}, \citenamefont {Wang}, \citenamefont {Gustavsson},\
  and\ \citenamefont {Oliver}}]{Kjaergaard2019}%
  \BibitemOpen
  \bibfield  {author} {\bibinfo {author} {\bibfnamefont {M.}~\bibnamefont
  {Kjaergaard}}, \bibinfo {author} {\bibfnamefont {M.~E.}\ \bibnamefont
  {Schwartz}}, \bibinfo {author} {\bibfnamefont {J.}~\bibnamefont
  {BraumÃŒller}}, \bibinfo {author} {\bibfnamefont {P.}~\bibnamefont
  {Krantz}}, \bibinfo {author} {\bibfnamefont {J.~I.-J.}\ \bibnamefont {Wang}},
  \bibinfo {author} {\bibfnamefont {S.}~\bibnamefont {Gustavsson}}, \ and\
  \bibinfo {author} {\bibfnamefont {W.~D.}\ \bibnamefont {Oliver}},\ }\href
  {\doibase 10.1146/annurev-conmatphys-031119-050605} {\bibfield  {journal}
  {\bibinfo  {journal} {Annual Review of Condensed Matter Physics}\ }\textbf
  {\bibinfo {volume} {11}},\ \bibinfo {pages} {null} (\bibinfo {year}
  {2020})}\BibitemShut {NoStop}%
\bibitem [{\citenamefont {Landig1}\ \emph {et~al.}(2019)\citenamefont
  {Landig1}, \citenamefont {Koski}, \citenamefont {Scarlino}, \citenamefont
  {M\"{u}ller}, \citenamefont {.~Abadillo-Uriel}, \citenamefont {Kratochwil},
  \citenamefont {Reichl}, \citenamefont {Wegscheider}, \citenamefont
  {Coppersmith}, \citenamefont {Friesen}, \citenamefont {Wallraff},
  \citenamefont {Ihn},\ and\ \citenamefont {Ensslin}}]{Landig2019}%
  \BibitemOpen
  \bibfield  {author} {\bibinfo {author} {\bibfnamefont {A.}~\bibnamefont
  {Landig1}}, \bibinfo {author} {\bibfnamefont {J.}~\bibnamefont {Koski}},
  \bibinfo {author} {\bibfnamefont {P.}~\bibnamefont {Scarlino}}, \bibinfo
  {author} {\bibfnamefont {C.}~\bibnamefont {M\"{u}ller}}, \bibinfo {author}
  {\bibfnamefont {J.}~\bibnamefont {.~Abadillo-Uriel}}, \bibinfo {author}
  {\bibfnamefont {B.}~\bibnamefont {Kratochwil}}, \bibinfo {author}
  {\bibfnamefont {C.}~\bibnamefont {Reichl}}, \bibinfo {author} {\bibfnamefont
  {W.}~\bibnamefont {Wegscheider}}, \bibinfo {author} {\bibfnamefont
  {S.}~\bibnamefont {Coppersmith}}, \bibinfo {author} {\bibfnamefont
  {M.}~\bibnamefont {Friesen}}, \bibinfo {author} {\bibfnamefont
  {A.}~\bibnamefont {Wallraff}}, \bibinfo {author} {\bibfnamefont
  {T.}~\bibnamefont {Ihn}}, \ and\ \bibinfo {author} {\bibfnamefont
  {K.}~\bibnamefont {Ensslin}},\ }\href {\doibase 10.1038/s41467-019-13000-z}
  {\bibfield  {journal} {\bibinfo  {journal} {Nature Communications}\ }\textbf
  {\bibinfo {volume} {10}},\ \bibinfo {pages} {5037} (\bibinfo {year}
  {2019})}\BibitemShut {NoStop}%
\bibitem [{\citenamefont {{Jaynes}}\ and\ \citenamefont
  {{Cummings}}(1963)}]{JCM1963}%
  \BibitemOpen
  \bibfield  {author} {\bibinfo {author} {\bibfnamefont {E.~T.}\ \bibnamefont
  {{Jaynes}}}\ and\ \bibinfo {author} {\bibfnamefont {F.~W.}\ \bibnamefont
  {{Cummings}}},\ }\href {\doibase 0.1109/PROC.1963.1664} {\bibfield  {journal}
  {\bibinfo  {journal} {Proceedings of the IEEE}\ }\textbf {\bibinfo {volume}
  {51}},\ \bibinfo {pages} {89} (\bibinfo {year} {1963})}\BibitemShut {NoStop}%
\bibitem [{\citenamefont {Gea-Banacloche}(1990)}]{Gea-Banacloche1990prl}%
  \BibitemOpen
  \bibfield  {author} {\bibinfo {author} {\bibfnamefont {J.}~\bibnamefont
  {Gea-Banacloche}},\ }\href {\doibase 10.1103/PhysRevLett.65.3385} {\bibfield
  {journal} {\bibinfo  {journal} {Phys. Rev. Lett.}\ }\textbf {\bibinfo
  {volume} {65}},\ \bibinfo {pages} {3385} (\bibinfo {year}
  {1990})}\BibitemShut {NoStop}%
\bibitem [{\citenamefont {Gea-Banacloche}(1991)}]{Gea-Banacloche1991pra}%
  \BibitemOpen
  \bibfield  {author} {\bibinfo {author} {\bibfnamefont {J.}~\bibnamefont
  {Gea-Banacloche}},\ }\href {\doibase 10.1103/PhysRevA.44.5913} {\bibfield
  {journal} {\bibinfo  {journal} {Phys. Rev. A}\ }\textbf {\bibinfo {volume}
  {44}},\ \bibinfo {pages} {5913} (\bibinfo {year} {1991})}\BibitemShut
  {NoStop}%
\bibitem [{\citenamefont {Retamal}\ \emph {et~al.}(1997)\citenamefont
  {Retamal}, \citenamefont {Saavedra}, \citenamefont {Klimov},\ and\
  \citenamefont {Chumakov}}]{Saavedra1997}%
  \BibitemOpen
  \bibfield  {author} {\bibinfo {author} {\bibfnamefont {J.~C.}\ \bibnamefont
  {Retamal}}, \bibinfo {author} {\bibfnamefont {C.}~\bibnamefont {Saavedra}},
  \bibinfo {author} {\bibfnamefont {A.~B.}\ \bibnamefont {Klimov}}, \ and\
  \bibinfo {author} {\bibfnamefont {S.~M.}\ \bibnamefont {Chumakov}},\ }\href
  {\doibase 10.1103/PhysRevA.55.2413} {\bibfield  {journal} {\bibinfo
  {journal} {Phys. Rev. A}\ }\textbf {\bibinfo {volume} {55}},\ \bibinfo
  {pages} {2413} (\bibinfo {year} {1997})}\BibitemShut {NoStop}%
\bibitem [{\citenamefont {Hermann-Avigliano}\ \emph {et~al.}(2015)\citenamefont
  {Hermann-Avigliano}, \citenamefont {Cisternas}, \citenamefont {Brune},
  \citenamefont {Raimond},\ and\ \citenamefont {Saavedra}}]{Hermann2015pra}%
  \BibitemOpen
  \bibfield  {author} {\bibinfo {author} {\bibfnamefont {C.}~\bibnamefont
  {Hermann-Avigliano}}, \bibinfo {author} {\bibfnamefont {N.}~\bibnamefont
  {Cisternas}}, \bibinfo {author} {\bibfnamefont {M.}~\bibnamefont {Brune}},
  \bibinfo {author} {\bibfnamefont {J.-M.}\ \bibnamefont {Raimond}}, \ and\
  \bibinfo {author} {\bibfnamefont {C.}~\bibnamefont {Saavedra}},\ }\href
  {\doibase 10.1103/PhysRevA.91.013815} {\bibfield  {journal} {\bibinfo
  {journal} {Phys. Rev. A}\ }\textbf {\bibinfo {volume} {91}},\ \bibinfo
  {pages} {013815} (\bibinfo {year} {2015})}\BibitemShut {NoStop}%
\bibitem [{SM()}]{SM}%
  \BibitemOpen
  \href@noop {} {\ }\bibinfo {note} {The Supplemental Material includes a video
  of the full dynamic of the Wigner function of the intracavity field, an
  analytical approach to the decoherence-free single atom problem, an analysis
  of the purity of the final state, and a study of the field dynamics in the
  Fock states basis, among other details.}\BibitemShut {Stop}%
\bibitem [{\citenamefont {de~Oliveira}\ \emph
  {et~al.}(1990{\natexlab{a}})\citenamefont {de~Oliveira}, \citenamefont {Kim},
  \citenamefont {Knight},\ and\ \citenamefont {Buek}}]{knight1990}%
  \BibitemOpen
  \bibfield  {author} {\bibinfo {author} {\bibfnamefont {F.~A.~M.}\
  \bibnamefont {de~Oliveira}}, \bibinfo {author} {\bibfnamefont {M.~S.}\
  \bibnamefont {Kim}}, \bibinfo {author} {\bibfnamefont {P.~L.}\ \bibnamefont
  {Knight}}, \ and\ \bibinfo {author} {\bibfnamefont {V.}~\bibnamefont
  {Buek}},\ }\href {\doibase 10.1103/PhysRevA.41.2645} {\bibfield  {journal}
  {\bibinfo  {journal} {Phys. Rev. A}\ }\textbf {\bibinfo {volume} {41}},\
  \bibinfo {pages} {2645} (\bibinfo {year} {1990}{\natexlab{a}})}\BibitemShut
  {NoStop}%
\bibitem [{\citenamefont {{Jozsa}}(1994)}]{1994JMO}%
  \BibitemOpen
  \bibfield  {author} {\bibinfo {author} {\bibfnamefont {R.}~\bibnamefont
  {{Jozsa}}},\ }\href {\doibase 10.1080/09500349414552171} {\bibfield
  {journal} {\bibinfo  {journal} {Journal of Modern Optics}\ }\textbf {\bibinfo
  {volume} {41}},\ \bibinfo {pages} {2315} (\bibinfo {year}
  {1994})}\BibitemShut {NoStop}%
\bibitem [{\citenamefont {Gilchrist}\ \emph {et~al.}(2005)\citenamefont
  {Gilchrist}, \citenamefont {Langford},\ and\ \citenamefont
  {Nielsen}}]{PhysRevA.71.062310}%
  \BibitemOpen
  \bibfield  {author} {\bibinfo {author} {\bibfnamefont {A.}~\bibnamefont
  {Gilchrist}}, \bibinfo {author} {\bibfnamefont {N.~K.}\ \bibnamefont
  {Langford}}, \ and\ \bibinfo {author} {\bibfnamefont {M.~A.}\ \bibnamefont
  {Nielsen}},\ }\href {\doibase 10.1103/PhysRevA.71.062310} {\bibfield
  {journal} {\bibinfo  {journal} {Phys. Rev. A}\ }\textbf {\bibinfo {volume}
  {71}},\ \bibinfo {pages} {062310} (\bibinfo {year} {2005})}\BibitemShut
  {NoStop}%
\bibitem [{\citenamefont {{Cl\'{e}ment Sayrin}}(2011)}]{SayrinThesis}%
  \BibitemOpen
  \bibfield  {author} {\bibinfo {author} {\bibnamefont {{Cl\'{e}ment
  Sayrin}}},\ }\href {https://tel.archives-ouvertes.fr/tel-00654082} {\bibfield
   {journal} {\bibinfo  {journal} {Ph.D. Thesis}\ } (\bibinfo {year}
  {2011})}\BibitemShut {NoStop}%
\bibitem [{\citenamefont {Penasa}\ \emph {et~al.}(2016)\citenamefont {Penasa},
  \citenamefont {Gerlich}, \citenamefont {Rybarczyk}, \citenamefont
  {M\'etillon}, \citenamefont {Brune}, \citenamefont {Raimond}, \citenamefont
  {Haroche}, \citenamefont {Davidovich},\ and\ \citenamefont
  {Dotsenko}}]{Penasa2016}%
  \BibitemOpen
  \bibfield  {author} {\bibinfo {author} {\bibfnamefont {M.}~\bibnamefont
  {Penasa}}, \bibinfo {author} {\bibfnamefont {S.}~\bibnamefont {Gerlich}},
  \bibinfo {author} {\bibfnamefont {T.}~\bibnamefont {Rybarczyk}}, \bibinfo
  {author} {\bibfnamefont {V.}~\bibnamefont {M\'etillon}}, \bibinfo {author}
  {\bibfnamefont {M.}~\bibnamefont {Brune}}, \bibinfo {author} {\bibfnamefont
  {J.~M.}\ \bibnamefont {Raimond}}, \bibinfo {author} {\bibfnamefont
  {S.}~\bibnamefont {Haroche}}, \bibinfo {author} {\bibfnamefont
  {L.}~\bibnamefont {Davidovich}}, \ and\ \bibinfo {author} {\bibfnamefont
  {I.}~\bibnamefont {Dotsenko}},\ }\href {\doibase 10.1103/PhysRevA.94.022313}
  {\bibfield  {journal} {\bibinfo  {journal} {Phys. Rev. A}\ }\textbf {\bibinfo
  {volume} {94}},\ \bibinfo {pages} {022313} (\bibinfo {year}
  {2016})}\BibitemShut {NoStop}%
\bibitem [{\citenamefont {Guerlin}\ \emph {et~al.}(2007)\citenamefont
  {Guerlin}, \citenamefont {Bernu}, \citenamefont {Deleglise}, \citenamefont
  {Sayrin}, \citenamefont {Gleyzes}, \citenamefont {Kuhr}, \citenamefont
  {Brune}, \citenamefont {Raimond},\ and\ \citenamefont
  {Haroche}}]{guerlin2007}%
  \BibitemOpen
  \bibfield  {author} {\bibinfo {author} {\bibfnamefont {C.}~\bibnamefont
  {Guerlin}}, \bibinfo {author} {\bibfnamefont {J.}~\bibnamefont {Bernu}},
  \bibinfo {author} {\bibfnamefont {S.}~\bibnamefont {Deleglise}}, \bibinfo
  {author} {\bibfnamefont {C.}~\bibnamefont {Sayrin}}, \bibinfo {author}
  {\bibfnamefont {S.}~\bibnamefont {Gleyzes}}, \bibinfo {author} {\bibfnamefont
  {S.}~\bibnamefont {Kuhr}}, \bibinfo {author} {\bibfnamefont {M.}~\bibnamefont
  {Brune}}, \bibinfo {author} {\bibfnamefont {J.-M.}\ \bibnamefont {Raimond}},
  \ and\ \bibinfo {author} {\bibfnamefont {S.}~\bibnamefont {Haroche}},\ }\href
  {\doibase 10.1038/nature06057} {\bibfield  {journal} {\bibinfo  {journal}
  {Nature}\ }\textbf {\bibinfo {volume} {448}},\ \bibinfo {pages} {889}
  (\bibinfo {year} {2007})}\BibitemShut {NoStop}%
\bibitem [{\citenamefont {Assemat}\ \emph {et~al.}(2019)\citenamefont
  {Assemat}, \citenamefont {Grosso}, \citenamefont {Signoles}, \citenamefont
  {Facon}, \citenamefont {Dotsenko}, \citenamefont {Haroche}, \citenamefont
  {Raimond}, \citenamefont {Brune},\ and\ \citenamefont
  {Gleyzes}}]{sebastien2019}%
  \BibitemOpen
  \bibfield  {author} {\bibinfo {author} {\bibfnamefont {F.}~\bibnamefont
  {Assemat}}, \bibinfo {author} {\bibfnamefont {D.}~\bibnamefont {Grosso}},
  \bibinfo {author} {\bibfnamefont {A.}~\bibnamefont {Signoles}}, \bibinfo
  {author} {\bibfnamefont {A.}~\bibnamefont {Facon}}, \bibinfo {author}
  {\bibfnamefont {I.}~\bibnamefont {Dotsenko}}, \bibinfo {author}
  {\bibfnamefont {S.}~\bibnamefont {Haroche}}, \bibinfo {author} {\bibfnamefont
  {J.~M.}\ \bibnamefont {Raimond}}, \bibinfo {author} {\bibfnamefont
  {M.}~\bibnamefont {Brune}}, \ and\ \bibinfo {author} {\bibfnamefont
  {S.}~\bibnamefont {Gleyzes}},\ }\href {\doibase
  10.1103/PhysRevLett.123.143605} {\bibfield  {journal} {\bibinfo  {journal}
  {Phys. Rev. Lett.}\ }\textbf {\bibinfo {volume} {123}},\ \bibinfo {pages}
  {143605} (\bibinfo {year} {2019})}\BibitemShut {NoStop}%
\bibitem [{\citenamefont {Albarelli}\ \emph {et~al.}(2016)\citenamefont
  {Albarelli}, \citenamefont {Ferraro}, \citenamefont {Paternostro},\ and\
  \citenamefont {Paris}}]{Albarelli2016}%
  \BibitemOpen
  \bibfield  {author} {\bibinfo {author} {\bibfnamefont {F.}~\bibnamefont
  {Albarelli}}, \bibinfo {author} {\bibfnamefont {A.}~\bibnamefont {Ferraro}},
  \bibinfo {author} {\bibfnamefont {M.}~\bibnamefont {Paternostro}}, \ and\
  \bibinfo {author} {\bibfnamefont {M.~G.~A.}\ \bibnamefont {Paris}},\ }\href
  {\doibase 10.1103/PhysRevA.93.032112} {\bibfield  {journal} {\bibinfo
  {journal} {Phys. Rev. A}\ }\textbf {\bibinfo {volume} {93}},\ \bibinfo
  {pages} {032112} (\bibinfo {year} {2016})}\BibitemShut {NoStop}%
\bibitem [{\citenamefont {Lewis-Swan}\ \emph {et~al.}(2020)\citenamefont
  {Lewis-Swan}, \citenamefont {Barberena}, \citenamefont {Muniz}, \citenamefont
  {Cline}, \citenamefont {Young}, \citenamefont {Thompson},\ and\ \citenamefont
  {Rey}}]{AnaMaria2019}%
  \BibitemOpen
  \bibfield  {author} {\bibinfo {author} {\bibfnamefont {R.~J.}\ \bibnamefont
  {Lewis-Swan}}, \bibinfo {author} {\bibfnamefont {D.}~\bibnamefont
  {Barberena}}, \bibinfo {author} {\bibfnamefont {J.~A.}\ \bibnamefont
  {Muniz}}, \bibinfo {author} {\bibfnamefont {J.~R.~K.}\ \bibnamefont {Cline}},
  \bibinfo {author} {\bibfnamefont {D.}~\bibnamefont {Young}}, \bibinfo
  {author} {\bibfnamefont {J.~K.}\ \bibnamefont {Thompson}}, \ and\ \bibinfo
  {author} {\bibfnamefont {A.~M.}\ \bibnamefont {Rey}},\ }\href {\doibase
  10.1103/PhysRevLett.124.193602} {\bibfield  {journal} {\bibinfo  {journal}
  {Phys. Rev. Lett.}\ }\textbf {\bibinfo {volume} {124}},\ \bibinfo {pages}
  {193602} (\bibinfo {year} {2020})}\BibitemShut {NoStop}%
\bibitem [{\citenamefont {de~Oliveira}\ \emph
  {et~al.}(1990{\natexlab{b}})\citenamefont {de~Oliveira}, \citenamefont {Kim},
  \citenamefont {Knight},\ and\ \citenamefont {Buek}}]{knight1990b}%
  \BibitemOpen
  \bibfield  {author} {\bibinfo {author} {\bibfnamefont {F.~A.~M.}\
  \bibnamefont {de~Oliveira}}, \bibinfo {author} {\bibfnamefont {M.~S.}\
  \bibnamefont {Kim}}, \bibinfo {author} {\bibfnamefont {P.~L.}\ \bibnamefont
  {Knight}}, \ and\ \bibinfo {author} {\bibfnamefont {V.}~\bibnamefont
  {Buek}},\ }\href {\doibase 10.1103/PhysRevA.41.2645} {\bibfield  {journal}
  {\bibinfo  {journal} {Phys. Rev. A}\ }\textbf {\bibinfo {volume} {41}},\
  \bibinfo {pages} {2645} (\bibinfo {year} {1990}{\natexlab{b}})}\BibitemShut
  {NoStop}%
\end{thebibliography}%

\clearpage

\begin{widetext}

\section{Supplementary Material: Deterministic Generation of Large Fock States}

\subsection{Fock states generation with a single two-level system}

The quasi-resonant interaction of a single two-level system with a coherent radiation field is well described by the Jaynes-Cummings Hamiltonian,
\begin{equation}
\label{SM_EQ:jcm}
H=\frac{\hbar\omega_0}{2}\sigma_z+\hbar\omega_c a^\dagger a + \hbar g (a\sigma_++ a^\dagger\sigma_-),
\end{equation}
where $\omega_0$ and $\omega_c$ are the two-level system and field frequencies respectively, $g=\Omega_0/2$ is the coupling frequency, $\hat{a}$ and $\hat{a}^\dagger$ are the field operators, and $\hat{\sigma}_+$ and $\hat{\sigma}_-$ raising and lowering operators of the two-level system. The evolution of the atom-field compound state $\rho$, is determined on resonance by the interaction Hamiltonian  $H_{\text{int}}=\hbar g (a\sigma_++ a^\dagger\sigma_-)$.

The atom-field compound state, which is initially factorizable, evolves into an entangled state. For a particular evolution time, this state get nearly disentangled and the field can be found in a mesoscopic superposition of coherent states (see Fig. 1 of the main text). After a given time, the field state evolves to one that closely resembles a Fock state, but slightly displaced in phase space. In order to show this analytically, we consider the two-level system initially in the excited state while the field is in a coherent state, $\ket{\psi(0)}=\ket{e}\ket{\alpha}$. The evolution of the total field-atomic state can be written within the JC solution as \cite{HarocheBook}:
\begin{equation}
\label{SM_EQ:sol_jcm}
	\ket{\psi(t)}=\sum_n C_n(\bar{n}) \ket{n} \left( \cos(gt\sqrt{n+1})\ket{e} -i\frac{\sqrt{n}}{\alpha}\sin(gt\sqrt{n})\ket{g} \right),
\end{equation}
where $\ket{e}$ and $\ket{g}$ are the excited and ground state of the two-level system, $\ket{n}$ is the field Fock state of $n$ photons, and $C_n$ are the probability amplitudes that represent the initial field state in the Fock basis. At time $t=0$, $\vert C_n(\bar{n})\vert^2=\bar{n}^ne^{-\bar{n}}/n!$ would be the probability to find $n$ photons in the initial coherent field with $|\alpha|^2=\bar{n}$.

This state can be displaced through a coherent displacement operator for the field $D(\beta)$ with amplitude $\beta$, so that
\begin{equation}
\ket{\varphi_\beta(t)}=D(\beta)\ket{\psi(t)}=\sum_n C_n(\bar{n}) D(\beta)\ket{n}\left(\dots \right)=\sum_m \ket{m}\otimes\ket{k_m(t)},
\end{equation}
where $\ket{m}$ is a Fock state of the field and $\ket{k_m(t)}$ is the two-level system state given by
\begin{equation}
\ket{k_m(t)}=\bra{m}\ket{\varphi_\beta(t)}=\mel{\psi(t)}{D(\beta)}{m}=\sum_n C_n(\bar{n}) \mel{n}{D(\beta)}{m}\left( \cos(gt\sqrt{n+1})\ket{e} -i\frac{\sqrt{n}}{\alpha}\sin(gt\sqrt{n})\ket{g} \right).
\label{eq_SM:vector_k_m}
\end{equation}

The density matrix of such a displaced state is
\begin{equation}
\rho(\beta,t)=\ket{\varphi_\beta(t)}\bra{\varphi_\beta(t)}=\sum_m\sum_l \ket{k_m(t)}\bra{k_l(t)}\otimes\ket{m}\bra{l}.
\end{equation}

By taking the two-level system partial trace, we can express the field density matrix as
\begin{equation}
\begin{split}
\rho_f(\beta,t)&=\sum_m\sum_l \Tr_{\text{at}}\left[\ket{k_m(t)}\bra{k_l(t))}\right]\ket{m}\bra{l};\\
&=\sum_m\sum_l \Tr_{\text{at}}[\bra{m}\ket{\varphi_\beta(t)}\bra{\varphi_\beta{
	}(t)}\ket{l}]\ket{m}\bra{l};\\
&=\sum_m\sum_l \Tr_{\text{at}}[\bra{m}{\rho(\beta,t)}\ket{l}]\ket{m}\bra{l};\\
&=\sum_m\sum_l F_{m,l}(\beta,t)\ket{m}\bra{l} ,
\end{split}
\end{equation}
where $F_{m,l}(\beta,t)=\Tr_{\text{at}}[\langle m \vert\rho(\beta,t)\vert\l\rangle]=\langle m \vert\rho_{f}(\beta,t)\vert\l\rangle$. This is a way to write the density matrix of the field in the Fock states basis. 

Notice that $F_{m,m}(\beta,t)$ is the fidelity of finding the field $\rho_f(\beta,t)$ in a Fock state $\ket{m}$. When the term $F_{m,m}(\beta,t)$ is maximized as a function of the displacement $\beta$ and evolution time $t$, the other diagonal elements are minimized in order to maintain the unit value of the trace. If the target Fock state is then $\ket{m}$, the function to maximize is
\begin{equation}
   F_{m,m}(\beta,t)=\sum_n\sum_{n'} C_n(\bar{n}) C_{n'}(\bar{n}) \mel{n}{D(\beta)}{m}\mel{n'}{D(\beta)}{m}^{*}\left(\cos(g\sqrt{n+1}t)\cos(g\sqrt{n'+1}t)+\frac{\sqrt{nn'}}{\vert\alpha\vert^2}\sin(g\sqrt{n}t)\sin(g\sqrt{n'}t)\right).
   \label{eq_SM:F_ml}
\end{equation}

To get a better sense of the function $F_{m,m}(\beta,t)$, we first analyse its terms \cite{knight1990b}
\begin{equation}
\mel{n}{D(\beta)}{m}=\mel{m}{e^{-\beta^2/2}\sum_{i=0}^{\infty}\frac{(\beta a^\dagger)^i}{i!}\sum_{j=0}^{\infty}\frac{(-\beta a)^j}{j!}}{n}
\label{eq.proj}
\end{equation}
with 
\begin{equation} (a^\dagger)^ia^j\ket{n}=\frac{\sqrt{n!}\sqrt{(n-j+i)!}}{(n-j)!}\ket{n-j+i}.
\end{equation}

Since we want to project this last term into a Fock-like state $\ket{m}$, we make $m=n-j+i$
\begin{description}
	\item[Case 1] $m\ge n$; $i=m-n+j$ with $j_{min}\in[0,n]$,
	\begin{equation}
	\begin{split}
\mel{n}{D(\beta)}{m}&=e^{-\beta^2/2}\beta^{m-n}\sqrt{n!}\sqrt{m!}\sum_{j=0}^{n}\frac{(-1)^j(\beta^2)^j}{(m-n+j)!j!(n-j)!}\\
	&=e^{-\beta^2/2}\beta^{m-n}\sqrt{\frac{n!}{m!}}L_n^{m-n}(\beta^2)
	\end{split}
	\label{eq_M}
	\end{equation}
	\item[Case 2] $m<n$; $j=n-m+i$ with $ i\in[0,m]$ 
	\begin{equation}
	\begin{split}
\mel{n}{D(\beta)}{m}&=e^{-\beta^2/2}(-\beta)^{n-m}\sqrt{n!}\sqrt{m!}\sum_{i=0}^{n}\frac{(-1)^{i}(\beta^2)^i}{(n-m+i)!i!(m-i)!}\\
	&=e^{-\beta^2/2}(-\beta)^{n-m}\sqrt{\frac{m!}{n!}}L_m^{n-m}(\beta^2)
	\end{split}
	\label{eq_N}
	\end{equation}
with $L_n^k(x)$ the Laguerre associated function
\begin{equation}
		L_n^k(x)=\sum_{r=0}^{n} \frac{(-1)^r (n+k)!}{(n-r)!r!(k+r)!}x^r.
	\end{equation}
\end{description}

We observe that Eq. (\ref{eq_SM:F_ml}) is difficult to approximate, not just because it is defined as a piecewise function, but because several terms of the summation in $n$ and $n'$ significantly contribute to the final result. The terms that contribute the most are those with $n$ and $n'\sim \bar{n}$, because the Poissonian coefficients $C_n(\bar{n})C_{n'}(\bar{n})$ are maximized when $n=n'=\bar{n}$, providing an envelope for the summation. However, the terms with $\mel{m}{D(\beta)}{n}$ can sharply vary amplitude as a function of $n$ for $|\beta|>0.15$, as the Laguerre polynomials can have several roots. The temporal evolution of the fidelity (term in parenthesis in Eq. (\ref{eq_SM:F_ml})) does not depend on the target state $\ket{m}$, but on the different terms of the summation $n$ and $n'$, and the initial average photon number $\bar{n}$. However, we numerically notice that the state with maximum fidelity are those with $\bar{n}=m$ (see Fig. 2 in the main text). Moreover, by inspecting numerical results, we find that the optimum time $\tau_{\text{F}}$, which maximizes the fidelity of the final state, is the time that maximizes the time dependent term of Eq. (\ref{eq_SM:F_ml}) when $n=n'=\bar{n}$, meaning $\left(\cos(g\sqrt{n+1}t)\cos(g\sqrt{n+1}t)+\sin(g\sqrt{n}t)\sin(g\sqrt{n}t)\right)$. The solutions of maximizing that term are given by
\begin{equation}
g\tau_{\text{F}}(n,l)=(2l+1)\frac{\pi}{2}(\sqrt{n+1}+\sqrt{n}),
\label{tiempos}
\end{equation}
where the integer $l$ represents the multiple solutions that periodically bring the field into a displaced Fock state. 

\subsection{Fock states generation with few two-level systems}

We compare the obtained analytical solution against numerical simulations for the optimum Fock state generation. Numerical errors in the field state evolution are negligible. To ensure this, we use a Hilbert space in the Fock basis that is much larger than the average number of photons of the initial coherent field. This is more than factor two larger for a large average photon number. We repeat the calculation varying the size of the Hilbert space to assess the convergence of the solution. In the case of a single two-level system interacting with the field, the numerical solutions agree with the analytic ones. Fig. \ref{fg_SM:gt_versus_n} shows the optimum times $g\tau_{\text{F}}$ to generate the highest fidelity Fock-like state of $n$ photons as a function of $n$, starting from a coherent state $\vert\alpha\vert^2=\bar{n}=n$ with its multiple solutions, without considering decoherence. We include here the information of how those times change when increasing the number of two-level system. We found that with a second two-level system, the times to generate a Fock state are twice as fast than the case with a single two-level system, and with a third two-level system it is three times faster. This is a consequence of the collective Rabi frequency, provided the right initial state (see Fig. 2 in the main text). Fock-like states are generated periodically during the evolution of the system, but every time with a slightly different fidelity. For different target states the optimum time $\tau_{\text{F}}$ can vary within this multiple repetition. This is shown in Fig. \ref{fg_SM:gt_versus_n}, where the best time to generate a target Fock-like state ``jumps" from one branch of the solution to another, as a consequence of maximizing the fidelity of the final state. \\

\begin{figure}[H]
\centering
  \includegraphics[width=0.8\textwidth]{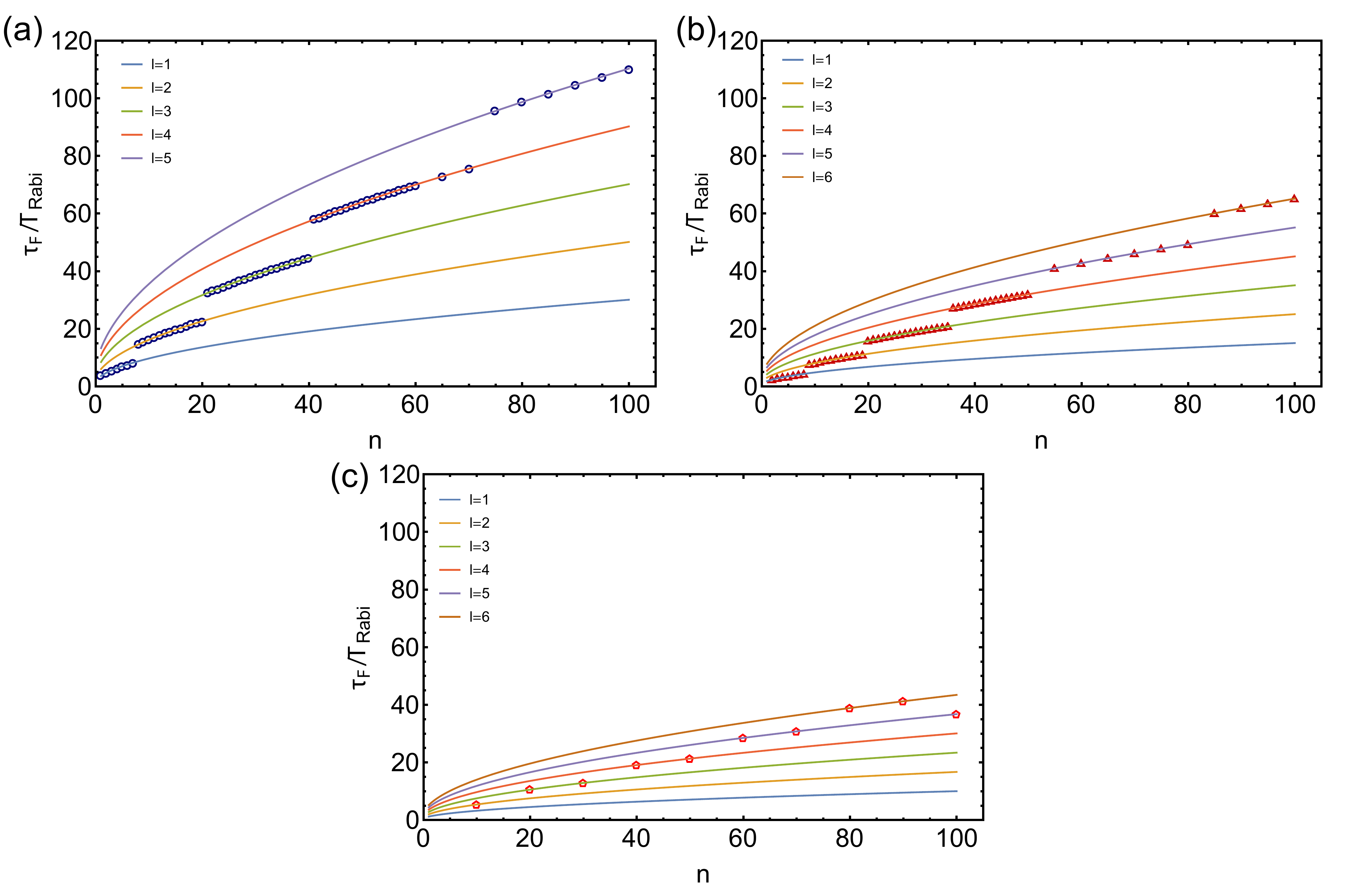}
 \caption{(a) Time $g\tau_{\text{F}}$ to generate the number states of $n$ photons with optimum fidelity. (a) shows the single two-level system initial state $\vert e\rangle$. (b) shows the two two-level systems initial state $(\vert ee\rangle + \vert gg\rangle)/\sqrt(2)$. (c) shows the three two-level systems initial state $(\vert eee\rangle+\vert egg\rangle+\vert geg\rangle+\vert gge\rangle)/2$. The Markers are numerical simulations. The solid lines correspond to the multiple analytic solutions for the case of a single two-level system, and in (b) and (c) they are scaled by the total number of two-level systems. }
 \label{fg_SM:gt_versus_n}
\end{figure}

Figure \ref{fg_SM:betas} shows the values of $\beta$ that generate the maximum fidelity Fock-like states, both for the one and two two-level systems cases. We were not able to find an analytical expression for the optimum displacement, however we give here the numerical fit for the single two-level system case, 
\begin{equation}
    \beta_{\text{F}}(l)\sim a(l)+b(l)g\tau_{\text{F}}(l),
    \label{Eq_SM:fit_betas}
\end{equation}
with $a(l)=0.13+1/(1.07+10.3\,l)$ when $l$ odd, $a(l)=-0.08-1/(2.65+2.6\,l)$ when $l$ even, $b(l)= (-1)^{l+1}\times[-0.0018+0.016/(-0.25+l)]$ for all $l$, where $l$ shows the explicit dependence on the multiple solutions for the evolution time. The displacements needed for the one and two two-level systems cases are similar, as Fig. \ref{fg_SM:betas} shows. The solid curves are the fitting curves found for the case of a single two-level system, showing the similarity with the case of two two-level systems. This evidences that the magnitude of displacements necessary to generate Fock-like states does not significantly change for increasing number of two-level systems.

\begin{figure}[h]
\centering
  \includegraphics[width=0.45\textwidth]{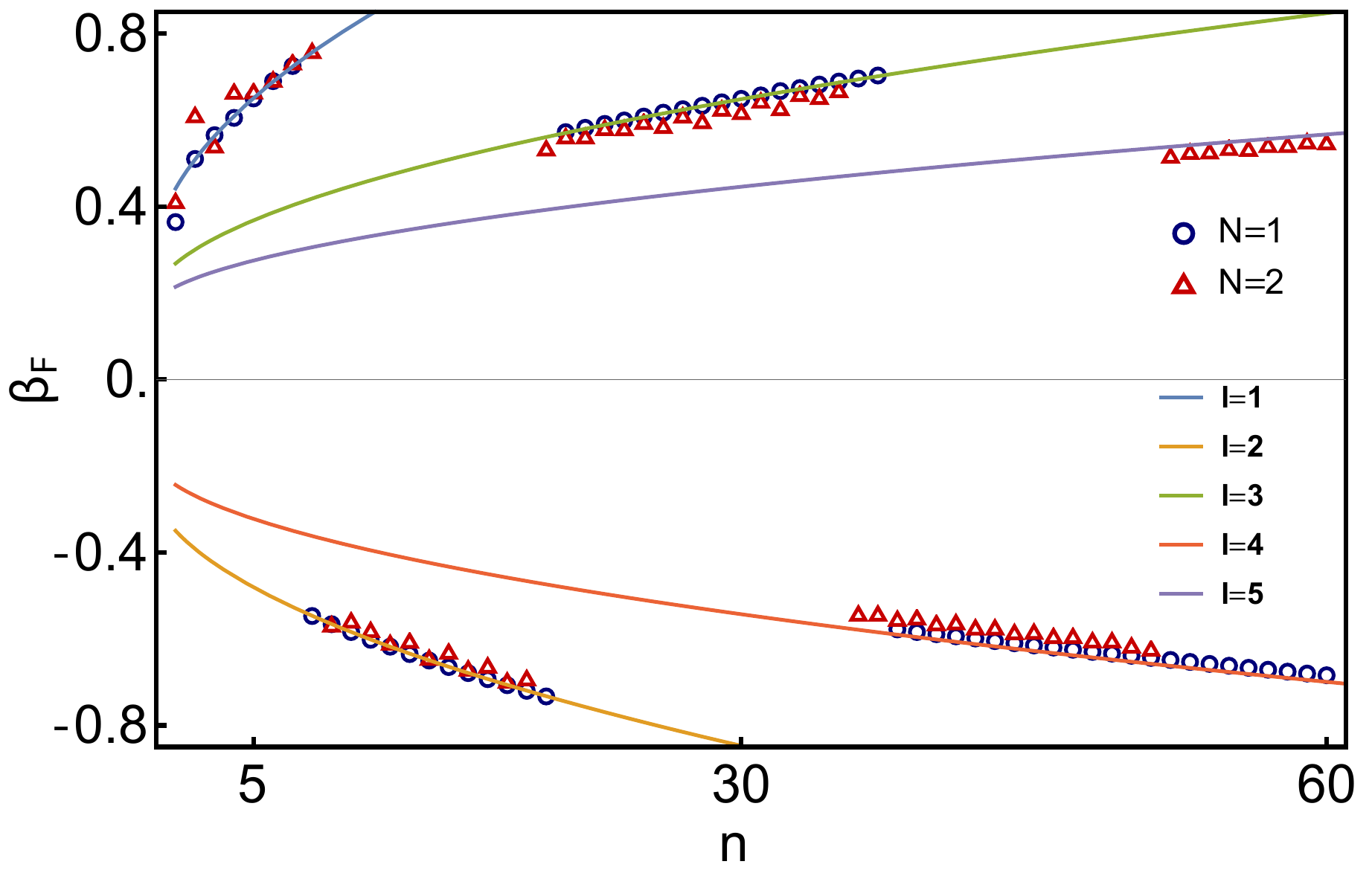}
 \caption{Displacement parameter $\beta$ that generate the best Fock-like state as a function of $n$, along with their fits (Eq. (\ref{Eq_SM:fit_betas})), for one (blue circles) and two two-level system (red triangles) case respectively.  Solid lines are fits to the single two-level system case.}
 \label{fg_SM:betas}
\end{figure}

In the following hyperlink, \url{https://www.dropbox.com/s/opjn6pnpmkbc3fd/A1N10.avi?dl=0}, we give a video showing the full dynamic of the Wigner function for the intracavity field for a target Fock state with $n=10$ photons. The video includes the first two $l$-solutions.

\subsection{Purity analysis}

We numerically analyze the purity and the fidelity of the generated Fock-like state after coherent displacement in phase space, $D(\beta)\rho_fD^{-1}(\beta)$. We compute the purity as
\begin{equation}
\begin{split}
\Tr[D(\beta)\rho_fD^{-1}(\beta)D(\beta)\rho_fD^{-1}(\beta)]&=\Tr[D(\beta)\rho_f\rho_fD^{-1}(\beta)]\\
&=\Tr[D^{-1}(\beta)D(\beta)\rho_f\rho_f]\\
&=\Tr[\rho_f\rho_f]=\Tr[\rho_f^2],
\end{split}
\end{equation}
which is independent of the applied displacement. Fig.~\ref{fg_SM:A_purity} shows the purity of the resulting state of the field as a function of $\vert\alpha\vert^2=\bar{n}=n$. This quantifies how much the two-level system(s) and the field states disentangled after the interaction. We observe that the purity in the single two-level system case is approximately constant with values near $80\%$. The purity drops below $70\%$ for the case of two two-level systems, suggesting that, even though more two-level systems can speed up the Fock state generation process, they will unavoidably compromise some of the coherence of the final state of the field.

\begin{figure}[h]
\centering
  \includegraphics[width=0.45\textwidth]{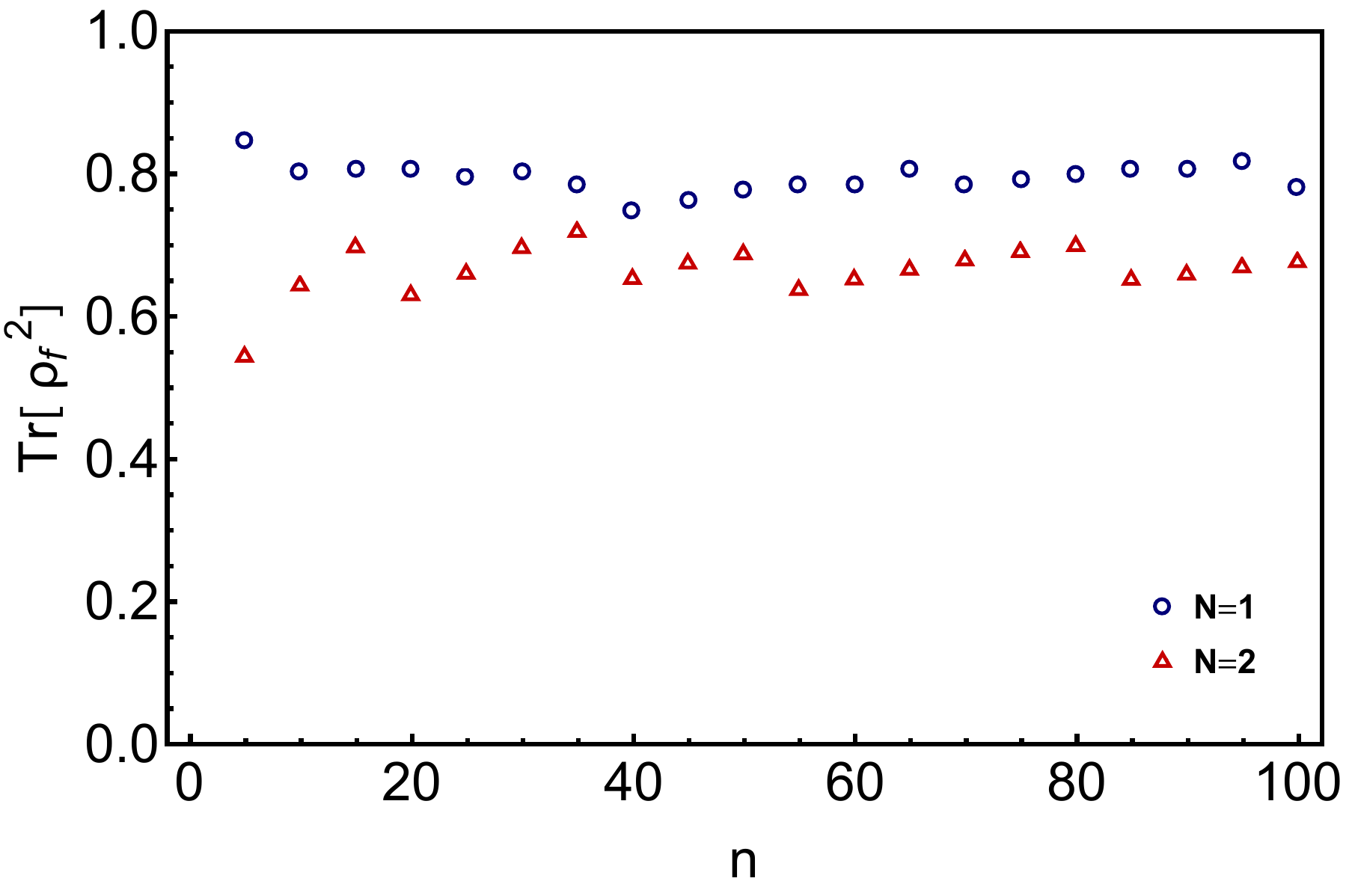}
 \caption{Purity of the final state of the field as a function of $n$, for the case of both one and two two-level systems.}
 \label{fg_SM:A_purity}
\end{figure}

\subsection{Density matrix analysis}

In order to get a deeper understanding of the dynamics of the field, we study the evolution of the trace elements of the field density matrix in the Fock state basis, 
\begin{equation}
	\rho_{nn}^f(t) = \Tr[\ket{\psi(t)}\bra{\psi(t)}\cdot (\ket{n}\bra{n}\otimes \mathbb{I})]
\end{equation}
where $\ket{\psi(t)}$ is the atomic-field state before displacement (see Eq. (\ref{SM_EQ:sol_jcm})). We color plot in Fig. \ref{fg_SM:evolucion_rho_nn}(a) the evolution of the diagonal elements of the field density matrix before the displacement for an initial coherent state with $\bar{n}=5$ photons (i.e. for a target Fock state of $n=5$). Vertical blue lines denote the times at which the Fock state is generated for different solutions of Eq. (\ref{tiempos}) denoted by $l$. This shows that the target Fock state is generated several times during the system evolution, but with different fidelities. In the case of low $n$, the best fidelity is found for $l=1$. Fig. \ref{fg_SM:evolucion_rho_nn}(b) and (c) show the probability distribution comparison of a theoretical Fock state displaced by $\beta_{\text{F}}$ (red curve) with the generated state (blue line, white dots), for $l=1$ and $l=2$ respectively.

\begin{figure}[h]
  \includegraphics[width=0.85\textwidth]{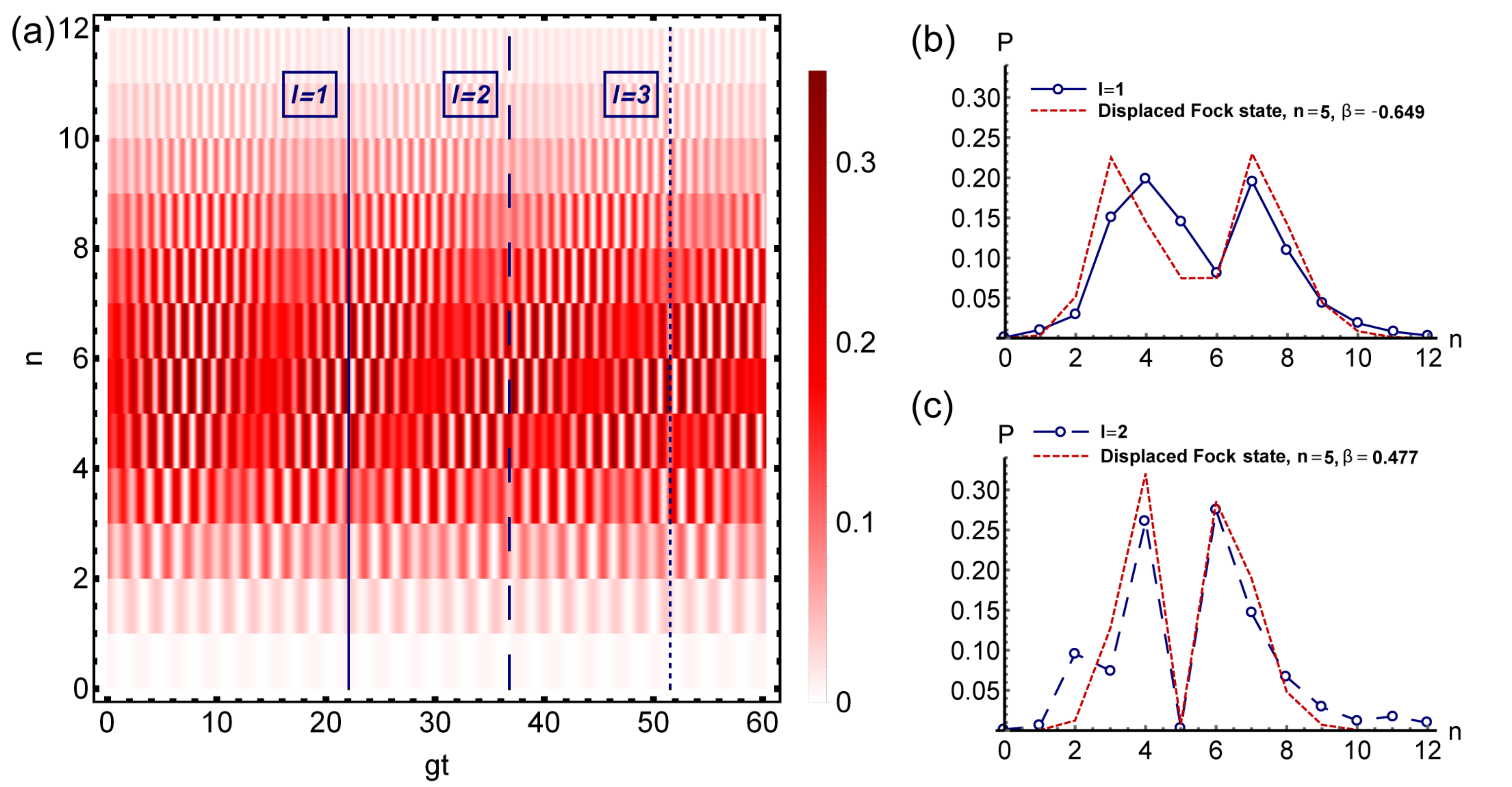}
 \caption{(a) Color plot of the trace elements of the field density matrix (before displacement) in the Fock basis for a target number state of $n=5$ photons, as a function of Fock states ($y$-axis) and interaction time $gt$ ($x$-axis). Vertical blue lines denote the integer $l$ that represents the multiple solutions that periodically bring the field into a displaced Fock-like state. (b) Probability distribution of the state for $l=1$ (blue curve) compared with an ideal displaced Fock state (red curve). (c) Probability distribution of the state for $l=2$ (blue curve) compared with an ideal displaced Fock state (red curve).}
 \label{fg_SM:evolucion_rho_nn}
\end{figure}

For completeness, in Fig. \ref{fg_SM:evolucion_rho_nn_displaced}(a) and (b) we plot how does the diagonal elements of the field density matrix look like when displaced for $\beta_\text{F}(l=1)$ and $\beta_\text{F}(l=2)$ respectively. In these cases, the emergence of a probability distribution that peaks at $n=5$ is evident. To highlight this, Fig. \ref{fg_SM:evolucion_rho_nn_displaced}(c) shows the probability distribution of both final states of the field.

\begin{figure}[h]
\centering
  \includegraphics[width=0.8\textwidth]{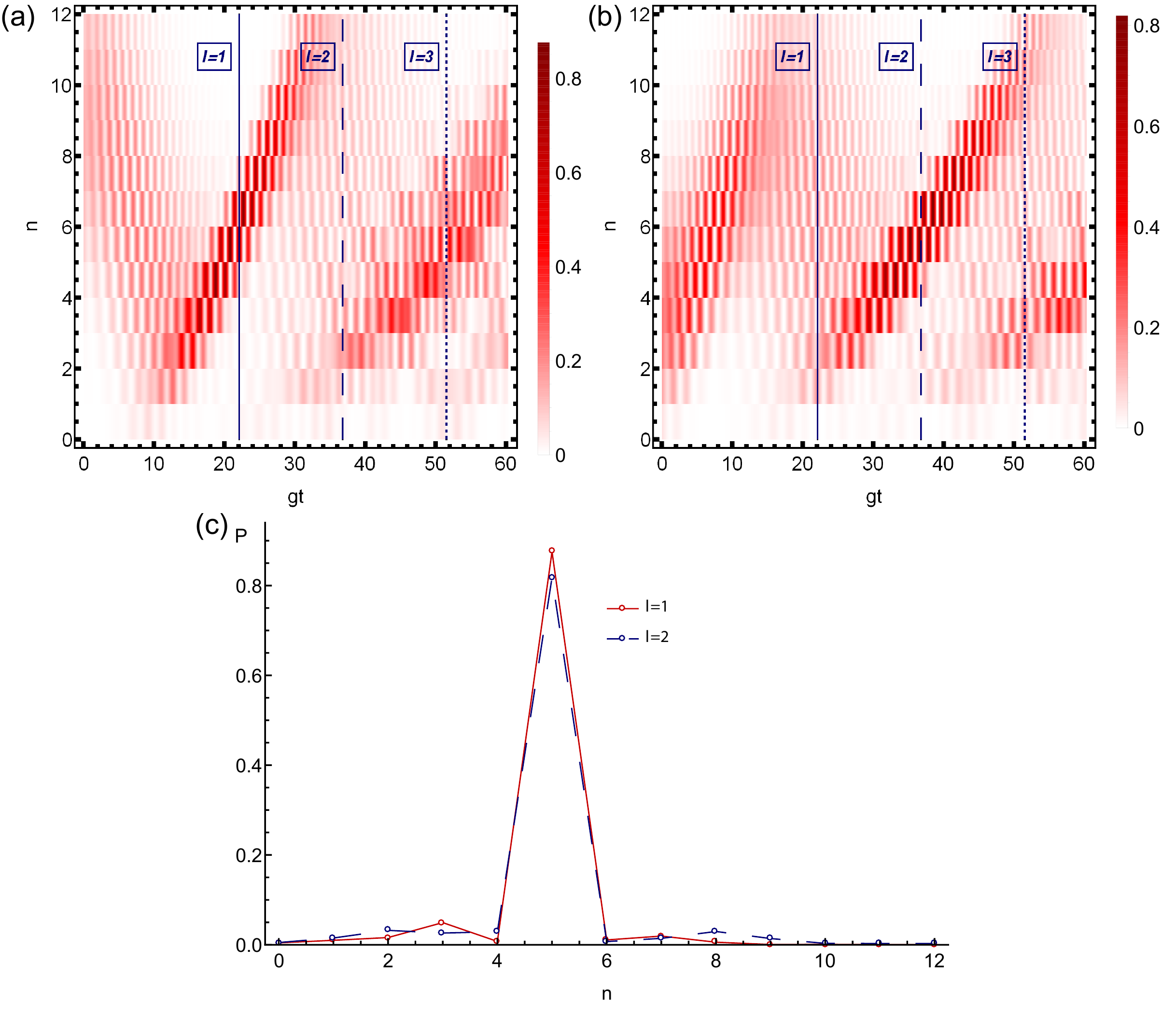}
 \caption{(a) Color plot of the trace elements of the field density matrix in the Fock basis after the field state is displaced by $\beta_\text{F}(l)$, for (a) $\beta_\text{F}(l=1)=0.649$ and (b) $\beta_\text{F}(l=2)=-0.477$. The two-level system is initially in the excited state and field starts in a coherent state with $\bar{n}=5$. We emphasize that these figures show the diagonal elements of the density matrices in the Fock after being displaced in phase space by the optimum value of $\beta_\text{F}(l)$. This is why it looks like the distribution starts near $n=10$ for (a) and near $n=3$ for (b).(c) shows the probability distribution for the states generated at $l=1$ (red solid curve) and the state at $l=2$ (blue dashed curve).}
 \label{fg_SM:evolucion_rho_nn_displaced}
\end{figure}

\subsection{Energy conservation and the role of coherent displacement}

In the absence of any two-level system decay or cavity decay, the energy of the system must be conserved at all times. The Jaynes-Cummings Hamiltonian describes the coherent energy exchange between the two-level system and the field, while conserving the total energy. Fig. \ref{fg_SM:energy}(a) shows this by displaying the dynamics of the expectation values for each energy contribution, given an initially excited two-level system and a coherent state with $|\alpha|^2=5$.

\begin{figure}[h]
\centering
  \includegraphics[width=0.45\textwidth]{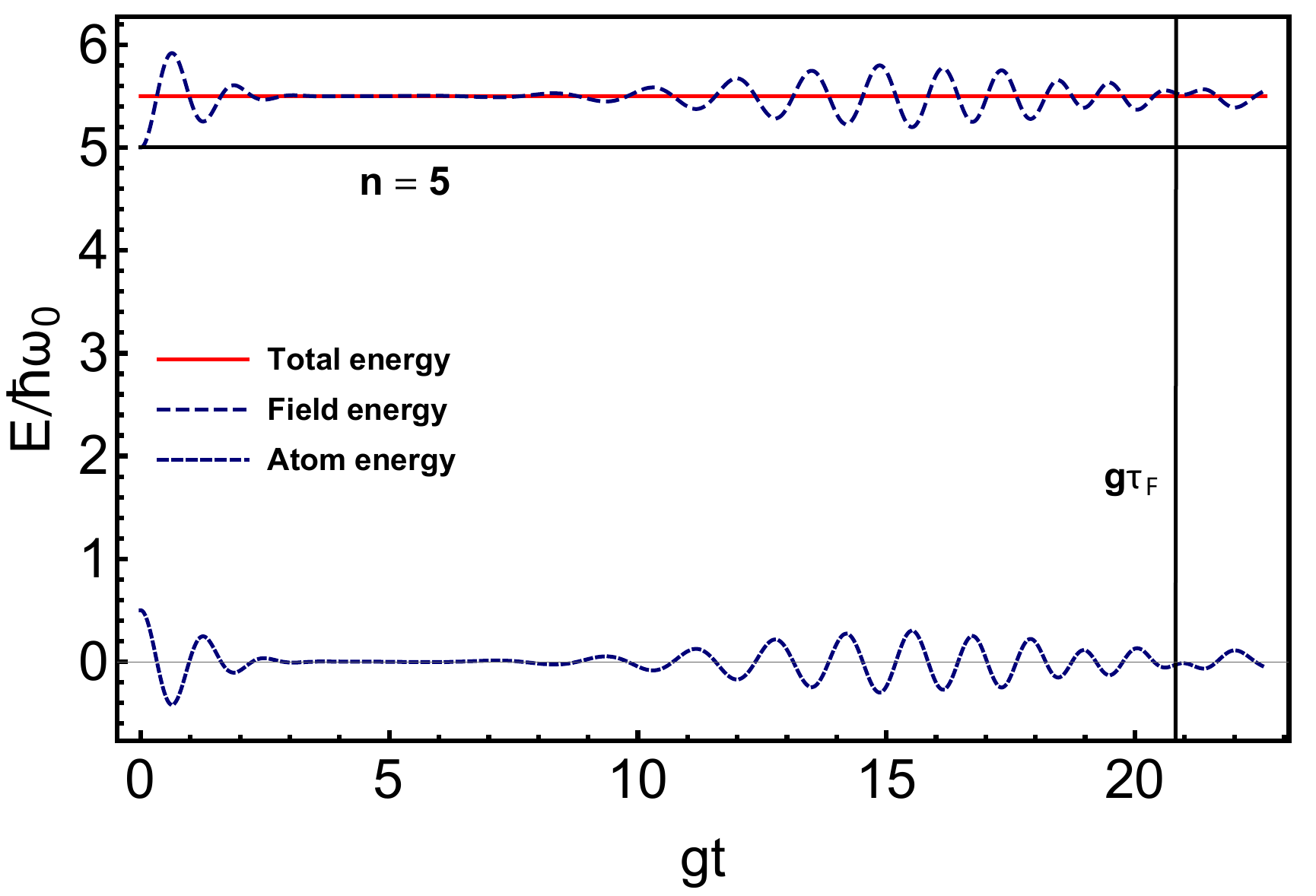}
 \caption{(a) Evolution of the expectation values of the energy for the complete system, two-level system and field as a function of $gt$. The field is initially in a coherent state of $|\alpha|^2=5$ and the two-level system is in the excited state. The target state is $\ket{n=5}$. The black vertical line denotes the time $\tau_{\text{F}}$ at which the Fock-like state with the largest fidelity is obtained. The black horizontal line denotes the energy of the target state. The energy difference between the field energy and the target state energy at a time $\tau_{\text{F}}$ is the one corrected by the coherent displacement $D(\beta_{\text{F}})$.}
 \label{fg_SM:energy}
\end{figure}

During the evolution of the system, the field can have an average energy above or below its initial value, depending on the initial state of the two-level system. More importantly, the final energy of the field can differ from the energy of the target Fock state. In the particular case of Fig. \ref{fg_SM:energy}(a) the energy of the field is always above the one of the target Fock state of $n=5$. This suggest that some energy should be substracted from the field at the end of the evolution (or added with the correct phase) to reach the target state. This is the role of the final coherent displacement in phase space, $D(\beta)$.

The mean energy of a Fock state $\ket{n}$ is $\hbar \omega n$. After displacing it by $D(\beta)$, it becomes $\hbar \omega (n+|\beta|^2)$. This is quite different from the case of displacing a coherent state $\ket{\alpha}$, where the average energy changes to $\hbar\omega|\alpha+\beta|^2$. Such energy change could be quite large due to interference effects. In other words, the displacement of a Fock state adds incoherently to the final mean energy of the field. 

In our proposal, we wait for the coherent state to evolve into a Fock-like state slightly displaced by $\beta_0$, whit energy $\hbar \omega (n+|\beta_0|^2)$, and at that time we stop the evolution and apply a displacement $D(\beta)$ to compensate for the energy exchanged with the two-level system. Such displacement changes the energy of the field to $\hbar \omega (n+|\beta_0-\beta|^2)$. If the displacements have the same magnitude but opposite phases ($\beta=-\beta_0$) we obtain a Fock state centered at the origin with the desired energy $\hbar \omega n$. This displacement occurs at the $\tau_{\text{F}}$ time denoted by a vertical dashed line in Fig. \ref{fg_SM:energy}(a).  

Since the two-level system can change its initial energy by at most one excitation, the energy difference between the final and initial field state is at most one photon, meaning that the necessary coherent displacement is $|\beta|\leq1$. In particular, we observe that the optimum Fock-like state is generated when the two-level system is left near a 50/50 superposition, meaning an energy of half a photon. This explains why the numerically obtained optimum displacements are always of the order of $|\beta_{\text{F}}|\sim\sqrt{0.5}$, as shown in Fig. \ref{fg_SM:betas}.

\subsection{Experimental robustness in CQED with Rydberg atoms}

We study the feasibility of the presented scheme in the context of CQED \cite{zhou2012PRL}. We consider the two most likely sources of error, which are the final coherent state displacement $\beta_{\text{F}}$ and the atom-cavity interaction time $g\tau_{\text{F}}$. In CQED the displacement $\beta$ is done by injecting a coherent microwave field from the side of the cavity, providing a small coupling into the cavity mode. The phase of the displacement can be manipulated relative to the phase of the coherent field initially injected in the cavity \cite{SayrinThesis}. On the other hand, the interaction time $\tau$ is set by the velocity of the atom sent through the cavity, where slower atoms means longer interaction times \cite{HarocheBook}.

\begin{figure}[h]
 \includegraphics[width=0.8\textwidth]{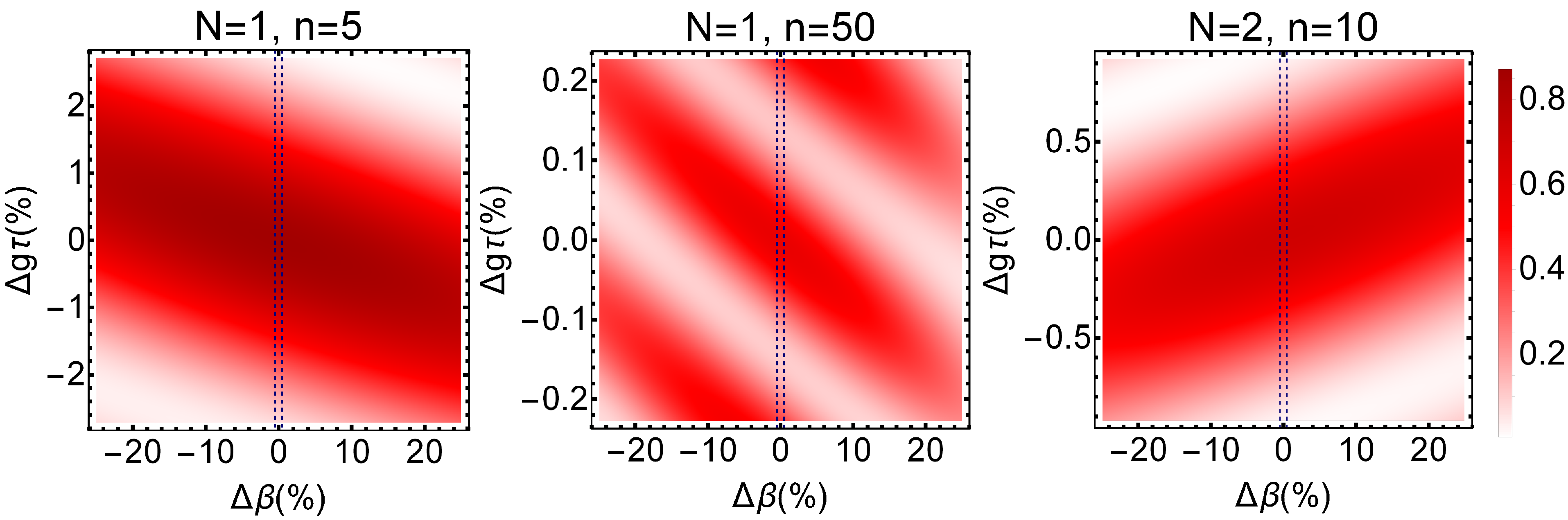}
 \caption{Fock state fidelity as a function of the coherent state displacement $\beta$ and the atoms-cavity interaction time in units of Rabi frequency $g\tau$, for the case of (a) $N=1$ with $n=5$, (b) $N=1$ with $n=50$, and (c) $N=2$ with $n=10$. The color scale represents the state fidelity with respect to the ideal number state. The vertical dashed lines represent reported experimental errors. Typical error in the interaction time are too small to be represented in these plots.}\label{fg:IV}
\end{figure}

Fig. \ref{fg:IV} shows the fidelity for the Fock state preparation  of  $n=5$ and   $n=50$ with a single two-level system, and of $n=10$ with two two-level systems for a range of values near $\beta_{\text{F}}$ and $\tau_{\text{F}}$, more than ten time larger than typical experimental errors \cite{SayrinThesis}. We observe that the most critical parameter to keep under control is the interaction time. However, the Fock state generation seems to be robust under typically reported experimental errors.

\end{widetext}
\end{document}